\documentclass[5p, times]{elsarticle}
\usepackage{amsmath,amssymb,amsfonts}
\usepackage{graphicx}
\usepackage{textcomp}
\usepackage{algcompatible}
\usepackage{algorithm}
\usepackage{pifont}
\usepackage{booktabs,makecell}
\usepackage{multirow}

\usepackage{siunitx}
\usepackage{fancyvrb}
\usepackage{verbatim}
\usepackage{svg}
\usepackage{xspace}
\usepackage[hyphens]{url}
\usepackage{hyperref}
\hypersetup{colorlinks=true,breaklinks=true}
\usepackage{listings}
\usepackage{numprint}
\usepackage{subcaption}
\usepackage{caption}
\usepackage{longtable}
\usepackage{threeparttable}
\usepackage[section]{placeins}
\usepackage{makecell}
\usepackage{pgfplots}
\usepackage{amsmath}
\usepackage{amssymb}
\usepackage{arydshln}
\usepackage{array}
\usepackage{tikz}
\usepackage{silence}
\usepackage{enumitem}
\usepackage{flushend}
\usepackage{numprint}
\usepackage{amssymb}
\usepackage{dblfloatfix}
\usepackage{amsmath}
\usepackage{soul}
\pgfplotsset{compat=1.15}

\definecolor{identical}{RGB}{10,148,10}
\definecolor{consistent}{RGB}{176,232,183}
\definecolor{exact}{RGB}{235,235,79}
\definecolor{plausible}{RGB}{255,166,0}
\definecolor{wellformed}{RGB}{217,120,120}
\definecolor{malformed}{RGB}{255,0,0}
\definecolor{noresult}{RGB}{169,169,169}
\definecolor{white}{RGB}{255,255,255}

\def\BibTeX{{\rm B\kern-.05em{\sc i\kern-.025em b}\kern-.08em
    T\kern-.1667em\lower.7ex\hbox{E}\kern-.125emX}}

\makeatletter
\def\BState{\State\hskip-\ALG@thistlm}
\makeatother

{}
\usepackage{lipsum}

\makeatletter
\newcommand\fs@betterruled{%
	\def\@fs@cfont{\bfseries}\let\@fs@capt\floatc@ruled
	\def\@fs@pre{\vspace*{5pt}\hrule height.8pt depth0pt \kern2pt}%
	\def\@fs@post{\kern2pt\hrule\relax}%
	\def\@fs@mid{\kern2pt\hrule\kern2pt}%
	\let\@fs@iftopcapt\iftrue}
\floatstyle{betterruled}
\restylefloat{algorithm}
\makeatother

\makeatletter
\renewcommand\st[1]{\@bsphack\@esphack}%
\makeatother

\usepackage{hyperref}
\hypersetup{pdfauthor={Md. Arafat Hossain}}

\makeatletter
\providecommand{\doi}[1]{%
  \begingroup
    \let\bibinfo\@secondoftwo
    \urlstyle{rm}%
    \href{http://dx.doi.org/#1}{%
      doi:\discretionary{}{}{}%
      \nolinkurl{#1}%
    }%
  \endgroup
}
\makeatother

\newcommand*\justify{%
	\fontdimen2\font=0.4em
	\fontdimen3\font=0.2em
	\fontdimen4\font=0.1em
	\fontdimen7\font=0.1em
	\hyphenchar\font=`\-
}

\newcommand{\eg}{\emph{e.g.}\xspace}

\newcommand{\ie}{\emph{i.e.}\xspace}

\newcommand{\dquotes}[1]{``#1''}
\newcolumntype{M}[1]{>{\centering\arraybackslash}m{#1}}

\newcommand\vartextvisiblespace[1][.7em]{%
  \makebox[#1]{%
    \kern.07em
    \vrule height.7ex
    \hrulefill
    \vrule height.7ex
    \kern.07em
  }
}

\newcommand{\messagetype}[1]{{\tt \justify{\textbf{#1}}}}
\newcolumntype{R}[1]{>{\raggedleft\let\newline\\\arraybackslash\hspace{0pt}}m{#1}}
\newcommand{\tracefont}[1]{\texttt{\footnotesize{\justify{#1}}}}
\newcommand{\tf}[1]{{\small \tt \justify{#1}}}
\newcolumntype{G}[1]{>{\centering\arraybackslash}p{#1}}

\usepackage{amssymb}

\journal{arXiv}

\begin{document}

\begin{frontmatter}



\title{Mining Service Behavior for Stateful Service Emulation}


\author[swin]{Md Arafat Hossain\corref{mycorrespondingauthor}}
\cortext[mycorrespondingauthor]{Corresponding authors}
\ead{mdarafathossain@swin.edu.au}


\author[swin]{Jun Han} 
\author[csu]{Muhammad Ashad Kabir\corref{mycorrespondingauthor}}
\ead{akabir@csu.edu.au}
\author[swin]{Steve Versteeg}
\author[monash]{Jean-Guy Schneider}
\author[unsw]{Jiaojiao Jiang}

\address[swin]{Department of Computing Technologies, Swinburne University of Technology, Hawthorn, VIC 3122, Australia}

\address[monash]{Department of Software Systems \& Cybersecurity, Monash University, Clayton, VIC 3800, Australia}

\address[csu]{School of Computing, Mathematics and Engineering, Charles Sturt University, Bathurst, NSW 2795, Australia}

\address[unsw]{School of Computer Science and Engineering, University of New South Wales, Sydney, NSW 2033, Australia}


\begin{abstract}
Enterprise software systems are increasingly integrating with diverse services to meet expanding business demands. Testing these highly interconnected systems presents a challenge due to the need for access to the connected services. Service virtualization has emerged as a widely used technique to derive service models from recorded interactions, for service response generation during system testing. Various methods have been proposed to emulate actual service behavior based on these interactions, but most fail to account for the service’s state, which reduces the accuracy of service emulation and the realism of the testing environment, especially when dealing with stateful services. This paper proposes an approach to deriving service models from service interactions, which enhance the accuracy of response generation by considering service state. This is achieved by uncovering contextual dependencies among interaction messages and analyzing the relationships between message data values. 
The approach is evaluated using interaction traces collected from both stateful and stateless services, and the results reveal notable enhancements in accuracy and efficiency over existing approaches in service response generation.
\end{abstract}




\end{frontmatter}



\section{Introduction}
\label{sec:introduction}
Enterprise software today is more connected than ever, linking many heterogeneous software services to support complex business processes and meet growing customer requirements. To effectively test such a system and ensure its quality, access to the connected services (as part of the testing environment) is required to fully exercise the system under test (SUT) in a realistic setting. However, such access for system testing is often restricted or even unavailable due to the cost, availability, or security concerns \cite{Michelsen:2012:SVR:2385447,du:2013}.

Traditionally, testing environments for enterprise software systems with externally connected services (or systems) are created by simulating or mocking the behavior of the connected services. Typical offerings include stubs \cite{hine:2016}, mock objects \cite{freeman2004mock} and fake frameworks \cite{Mackinnon:2001}. Virtual machine provides another option \cite{sugerman2001virtualizing, li2010selecting}, replicating the physical machine where the connected services are deployed. Containerization provides yet another way to create lightweight, executable hosting environments for such services \cite{docker, openvz, lmctfy}. Despite those strategies, achieving a testing environment that is \emph{scalable} and \emph{practical} for business enterprise software systems remains a challenge.

\textit{Service virtualization} has gained popularity as an approach to creating testing environments without the need to access the actual connected services. Instead, it uses executable service models in their place~\cite{Michelsen:2012:SVR:2385447}. The \emph{executable service models} (or \emph{virtual services}) are created and deployed in the testing environment to emulate the actual services for interacting with the system under test (SUT). These virtual services act as stand-ins, accepting requests from the SUT and generating appropriate replies, all while maintaining expected performance.
The use of such executable service models (in the place of actual services) presents the possibility of achieving realistic and yet scalable testing environments \cite{hine:2016}. 

The creation of virtual services can be done either manually \cite{hine:2016} 
or automatically \cite{du:2013, du:2015interaction, versteeg2016enhanced, versteeg:2016, enicser2018testing}. 
Manually defining service models requires knowledge of the underlying services, which is not always available, especially for legacy services or systems \cite{ghosh:1999}. Furthermore, it is very time-consuming and error-prone \cite{versteeg:2016}. 
On the other hand, a number of efforts have explored the use of automatically generated service models from service interaction traces,   
which requires no explicit knowledge of the underlying services and their message structures. They include
\textit{Opaque Service Virtualization (OSV)} \cite{versteeg:2016} and \textit{Sequence-to-Sequence Based Virtualization (SSBV)}, and have achieved different degrees of success.

Software services can be classified as either stateless or stateful. In stateless services, the response to a request is independent of any previous requests or operations \cite{su:2005slingshot}. For example, the response messages (HTTP response) from an online newspaper do not depend on the preceding requests and a reader can send HTTP requests to connect with the server in a stateless manner.
In contrast, a stateful service relies on the context provided by previous requests \cite{su:2005slingshot}. An example of such a service is a banking service, where an account has to be opened before receiving any financial service, the response to a withdrawal request depends on the requested withdrawal amount and the current balance of that account, and every change made to the account is considered as a change of state for that account. Therefore, the response to a request from such a stateful service depends on the preceding requests (or its internal state).
Existing service virtualization techniques have achieved a degree of effectiveness only for stateless services \cite{du:2013, du:2015interaction, versteeg2016enhanced, versteeg:2016} as a response is synthesized only based on the incoming request message without considering the service state or prior service interactions.  
Only a very limited attempt for service virtualization has been made for stateful services~\cite{enicser2018testing}. 

In this paper, we present a new method for extracting behavior models from interaction traces of both stateful and stateless services, and utilizing these models to generate precise response messages for incoming requests at system runtime. A mined service behavior model captures the dependencies between a service's messages (as inferred or dictated by the service state), determines and formulates the appropriate response message for each incoming request.
Based on our previous work in identifying message types and message formats \cite{hossain2022extracting}, this paper makes the following major contributions
\footnote{This article is based on the research first written in a PhD thesis~\cite{thesisArafat}. Also note that the outcomes of this research have been used in a consequent study reported in~\cite{hossain2022:datamodel} (with a new focus on the discovery and use of a relational data model for message payloads in service response generation), where only an overview of this research was included while referring to the PhD thesis for details and this research was not claimed as new contribution in~\cite{hossain2022:datamodel}. After further refinement, this article formally publishes the PhD research with the detailed approach, algorithms, and evaluation results.}:
\begin{itemize}
  \item \emph{Message dependency model:}
  A message dependency model is introduced to capture the dependency relationships between the messages of a service as embedded in its interaction trace. These dependencies reflect the service state changes and how these changes determine the types of response messages for incoming request messages. The model also recognizes that the service may be in different states for different application data records (the service maintains) at any given time.
  \item \emph{Response generation:} 
A response generation process is formulated to derive a more precise response message for any incoming request, at runtime. 
More specifically, the type of response message is identified by the message dependency model, the message structure is determined by the message format, and the message payload is populated from the service interaction traces and according to payload relationships. 
 
\end{itemize}
Our approach has been applied to a diverse set of real-world software services, including stateful and stateless services, where their behavior models are derived from their interaction traces and then used to generate responses for incoming requests (from systems under test) at runtime. 
The experimental results have shown that our approach achieves up to 50\% improvements in the accuracy of the generated service responses, compared to existing approaches.
In addition, our approach is also extremely efficient, generating responses in a fraction of a millisecond, which is not only faster than existing methods but also quicker than the actual services themselves.

\smallskip

The rest of this paper is organized as follows: Section \ref{sec:motivation} motivates our work using a typical example service with a sample interaction trace. Section \ref{sec:relatedwork} discusses the related work and highlights the research gap to be filled. Section \ref{sec:approach} presents the detailed techniques of our approach, including those for service model derivation and response message generation. Section \ref{sec:probabilisticModel} describes the techniques for inferring (alternative) probabilistic message dependency models (in contrast to the deterministic models of Section~\ref{sec:approach}). Section \ref{sec:evaluation} reports the experimental results while Section \ref{sec:discussion} 
discusses the limitations of the proposed approach. Finally, Section \ref{sec:conclusion} provides some concluding remarks.



\section{Background and Problem Analysis}
\label{sec:motivation}

\begin{figure}[!t]
\centering
\begin{subfigure}{0.4\textwidth}
    \includegraphics[width=\textwidth]{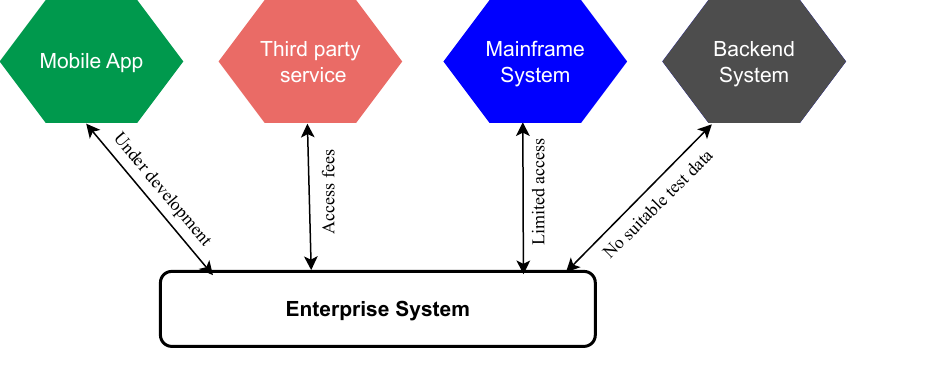}
    \caption{Actual operating environment}
    \label{fig:actualoperation}
\end{subfigure}
\begin{subfigure}{0.4\textwidth}
    \includegraphics[width=\textwidth]{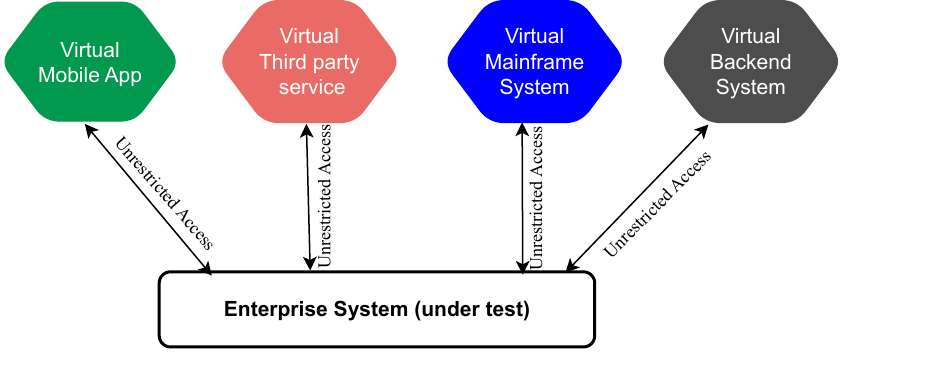}
    \caption{Virtualization-based testing environment} 
    \label{fig:virtualizationtestingenv}
\end{subfigure} 
 \caption{Enterprise system in actual and virtualized testing environments}
 \label{fig:suttestingenvironment}
\end{figure}

This section analyzes a typical enterprise system environment to highlight the research problems to be investigated.
Fig.~\ref{fig:actualoperation} shows an enterprise system connected with various other software services in its operating environment, including a mobile app, a third-party service, a mainframe system, and another backend system. Testing the enterprise system necessitates access to these connected services and systems. However, such access may not always be possible due to such constraints as the costs and availability of the connected services and systems. Service virtualization~\cite{Michelsen:2012:SVR:2385447} has been proposed as an alternative to testing the system, in such a way that does not require access to the actual connected services and systems. 

Fig.~\ref{fig:virtualizationtestingenv}
depicts the enterprise system within a testing environment that utilizes service virtualization. In this setup, the system under test (SUT) interacts with ``virtual" services that replicate the specific behaviors of the corresponding real services and systems. For example, the  \emph{Mobile App} in the actual operating environment (Fig.~\ref{fig:actualoperation}) is replaced by the \emph{Virtual Mobile App} in the testing environment (Fig.~\ref{fig:actualoperation}), where the \emph{Virtual Mobile App} emulates the behavior of the actual \emph{Mobile App}. 
A virtual service is based on a model of the corresponding actual service, and it emulates the behavior of the actual service by 
generating responses with proper payloads upon receiving requests from the SUT. Our main goal in this research is to derive an executable model for a service by mining its interaction trace and then use the model for response generation during its runtime interaction with the SUT. The key question to be answered is how close the derived service model's behavior is to the actual service's behavior.

In general, a response from a service depends not only on the request received but also on the service state, which is particularly important for a stateful service. For example, a search request from an identity management system (an enterprise system) on a staff directory service should not only lead to a search response but also with the corresponding staff member's information. A non-search response (\eg, a delete response) or a wrong type of search response (\eg, \tracefont{staff not found} instead of the expected staff information) will probably confuse the calling enterprise system and cause unexpected behavior such as premature termination. Furthermore, a response with another staff member's information (inaccurate payload), such as the supervisor's information instead of a trainee's information, may not cause premature system termination but may result in unauthorized access to sensitive information. In a virtualization-based testing environment, therefore, it is critically important that a virtual service generates accurate responses for incoming requests from the enterprise system under test. 

\begin{table}[!t]
    \centering
    \caption{Simplified interaction trace}
    \label{tab:exampleinteractionlibrary}
    \includegraphics[width=.48\textwidth]{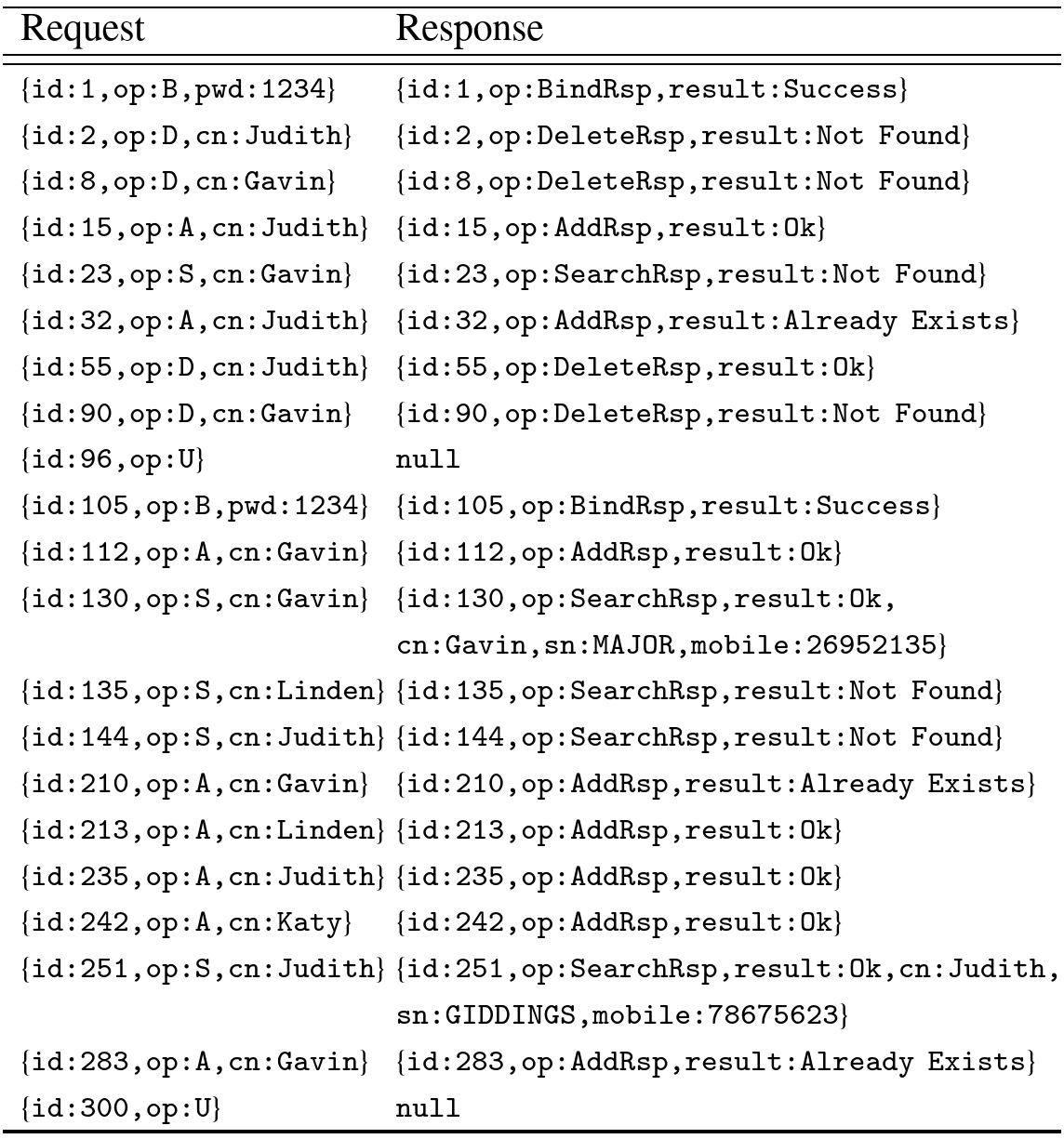}
\end{table}

To further examine the issue of service interaction behavior, let us consider a specific enterprise system scenario. The CA Identity Manager (CA IM) \cite{caidentitymanager} is an enterprise-grade identity management system that uses a directory service to manage the digital identities of personnel in an organization and to control access to its enterprise resources. The directory service and its interactions with CA IM follow the Light Weight Directory Access Protocol (LDAP) \cite{yeong:1995}. In a virtualization-based testing environment for this scenario, the CA IM is the system under test (SUT), and the personnel directory is a service in the testing environment.    
Table~\ref{tab:exampleinteractionlibrary} displays a portion of an example interaction trace between CA IM and the personnel directory.\footnote{An interaction trace is a collection of service interactions, where each interaction is a pair of request and response messages.} in a simplified form. 
It is important to note that for presentation clarity in this paper, a simplified representation of LDAP messages is used; however, the actual techniques and tools are intended to work with original LDAP messages, ensuring their actual formats are preserved.

The example interaction trace shown in Table \ref{tab:exampleinteractionlibrary} includes five types of request messages: \messagetype{Bind}, \messagetype{Add}, \messagetype{Search}, \messagetype{Delete}, and \messagetype{Unbind}, which are represented in the trace as \messagetype{op:B}, \messagetype{op:A}, \messagetype{op:S}, \messagetype{op:D}, and \messagetype{op:U}, respectively. Each type of request generates a corresponding response, namely \messagetype{op:BindRsp}, \messagetype{op: AddRsp}, \messagetype{op:SearchRsp}, and \messagetype{op:DeleteRsp}, with the \messagetype{Unbind} request returning no response (\ie, \textit{null})\footnote{An LDAP server does not send any response for an \messagetype{Unbind} request because the client intends to disconnect from the server.}. The response result codes include \messagetype{result:Ok}, \messagetype{result:Not found}, and 
\messagetype{result:AlreadyExists}. In addition, the \messagetype{cn} (common name) field uniquely identifies an entry managed by the service. For the purposes of this paper, this field is referred to as the \textit{key field}, and its value is called the \textit{key payload}. In practice, the \messagetype{dn} (distinguished name) field identifies an entry or record in the LDAP directory, which contains a \messagetype{cn}, at least one \messagetype{ou} (organizational unit), and one or more \messagetype{dc}'s (domain components). This simplification is made for the simplicity of discussion. Similarly, for other protocols/ or services, the value of the key attribute is considered as the \textit{key payload}.

As shown in Table~\ref{tab:exampleinteractionlibrary}, identical types of requests with different \textit{key payloads} result in distinct response types. For example, a \messagetype{Delete} request with {\small \tt `id:55'} leads to a \messagetype{DeleteRsp} with the response code \messagetype{result:Ok}. In contrast, a \messagetype{Delete} request with {\small \tt `id:90'} has a \messagetype{DeleteRsp} with the response code \messagetype{result:Not found}. This indicates that the service's state differs for the key payloads \messagetype{Judith} and \messagetype{Gavin}.
Moreover, performing the same type of request on the same record at different times can yield different responses. For instance, a \messagetype{Delete} request with {\small \tt `id:2'} results in a \messagetype{DeleteRsp} with the response code \messagetype{result:Not found}, while a \messagetype{Delete} request with {\small \tt `id:55'} has a \messagetype{DeleteRsp} with a different response code \messagetype{result:Ok}, indicating that the service's state for the key payload \messagetype{Judith} has changed between the two requests.
As detailed in Table \ref{tab:exampleinteractionlibrary}, an \messagetype{Add} request on the same record with {\small \tt `id:15'} was made between the two \messagetype{Delete} requests, modifying the service's state for that record. This highlights that the state of a service can be altered by executing certain requests, which, in turn, impacts the responses to subsequent requests.

Let us further explore the payload fields of messages, which can be categorized into: (i) protocol meta-data and (ii) application data. Protocol meta-data is used to identify or describe the message itself and may include correlation information between messages. 
For example, LDAP messages has an \messagetype{id} field as a protocol-meta data field
(see Table \ref{tab:exampleinteractionlibrary}). 
The \emph{message id} meta-data field can also be found in the POP3, SMTP, SMB, and LDAP protocols.
On the other hand, the ICMP, TCP/IP, and HTTP protocols have the \emph{message length} meta-data field in their messages, and the NTP, ICMP, and PTP protocols use \emph{timestamp} in their messages as a meta-data field. Therefore, the messages of any service that uses these protocols contain some protocol-meta data field(s) as well as application data field(s). 
In contrast, the application data fields of a message are generated and maintained by the service, manifesting as message payloads and usually related to a specific application data entity or record. For example, the response message with {\small \tt `id:130'}  in Table~\ref{tab:exampleinteractionlibrary} includes a \messagetype{mobile number}, a \messagetype{surname}, and the key payload {\messagetype{Gavin}}, all of which constitute application data pertaining to {\messagetype{Gavin}}'s record. 

It can be further observed that there are some simple relationships between the message payload fields, in general or for specific message types. For example, in LDAP messages (Table~\ref{tab:exampleinteractionlibrary}), the \emph{message id} fields are the same for every request and response message pair, and so are the \messagetype{cn} fields between the successful \messagetype{Search} request and response messages.

The following observations can be made from the above analysis:
\begin{enumerate}
    \item The response type and payloads of a stateful service to a request not only depend on the incoming request but also on the service's state.
    \item The service's state is the result of and may be changed by, prior requests.
    \item The service's state may be specific to particular application data records that the service maintains, and the service may be in different states for different data records at any given time.
   \item There may exist specific relationships (\eg, equality) between payloads or message fields of the same or different messages in a service's interactions.
\end{enumerate}
In this paper, our goal is to
(1) identify the dependencies between message types and between message types and message data fields (\ie, application data records)
by mining its interaction trace, thereby, forming an accurate behavior model of the service, and 
(2) use this service model (with some existing payload population techniques~\cite{du:2013}) 
to synthesize more precise responses for service requests (\ie, serving as the corresponding virtual service in a testing environment). Note that in this section, some terminology is introduced when discussing the sample LDAP message trace (such as \emph{interaction}, \emph{(application) data entry or record}, and \emph{key payload}), and they are equally applicable to service traces using other messaging protocols.



\section{Related Work}
\label{sec:relatedwork}
A suitable testing environment is essential for evaluating a highly connected enterprise system. Several techniques can be employed to build these environments, such as emulating the particular behaviors of connected services, creating replicas of physical devices or machines, or deducing service models from the existing services. This section presents an overview of traditional testing environments and reviews existing research in service virtualization. Current approaches to inferring control models of systems from their interaction traces are also explored, which can aid in simulating stateful services when integrated with service virtualization.

\subsection{Traditional Techniques for Creating Testing Environments}
\label{subsec:traditionaltestingtechniques}
HORUS \cite{gibbons:1987stub} is a technique for generating software stubs that marshal and unmarshal data values in multi-language Remote Procedure Calls (RPC). Mock objects \cite{freeman2004mock} and fake \cite{Mackinnon:2001} simulate the behavior of complex real objects in software systems and provide support for running unit tests, especially when it is impractical or impossible to access the actual objects. A number of other mocking techniques are also available, including jMock \cite{freeman2004mock}, EasyMock \cite{freese:2002easymock} and DynaMock \cite{massol:2003junit}, to mimic certain behavior of objects during testing. However, mocking objects requires intimate knowledge of the objects to be simulated as they require the re-implementation of some of their internal parts for each testing scenario. 

Another group of techniques provide  testing environments by replicating physical devices and running multiple virtual machines simultaneously. Hardware virtualization tools, such as VirtualBox \cite{li2010selecting}, VMWare \cite{sugerman2001virtualizing} and Plex86 \cite{plex86},  facilitate the testing of interconnected systems by replicating and deploying the connected services on virtual machines, providing an isolated environment for their users. However, these virtualization techniques 
suffer scalability problems as they replicate the hardware and software environments \cite{sanchezsqueezing}. 

Containerization is another technology of providing isolated runtime environments for applications and services, which involves bundling an application
with its libraries, configuration files, and dependencies (\eg, connected services).
Examples include Docker \cite{docker}, Openvz \cite{openvz} and Google lmctfy \cite{lmctfy}.  While providing improvements, containerization still suffers the scalability problem \cite{sanchezsqueezing} as a container still has a (lightweight) version of the application's operating environment, which runs inside the host system. 

\begin{table}[!t]
    \centering
    \caption{Summary of related work in service virtualization. The symbol $\sim$ indicates partial support, $\times$ indicates no support, and \checkmark indicates full support.}
    \label{tab:relatedwork}
    \begin{center}
    \resizebox{\linewidth}{!}{
\begin{tabular}{ l  c  c  c  } 
 \hline
 Approach & Automatic & Statefulness & Message Format \\ 
 \hline\hline
 CA SV \cite{caservicevirtualization} & $\sim$ & $\times$ & \checkmark \\ 
 
 ServiceV Pro \cite{serviceVPro}  & $\sim$ & $\times$ & \checkmark\\ 
 
  IBM SV \cite{IBMservicevirtualization}  & $\sim$ & $\times$ & \checkmark \\ 
 
 WireMock \cite{wiremock} & $\sim$ & $\times$ & \checkmark \\ 
 
 HoverFly \cite{hoverfly}  & $\sim$ & $\times$ & \checkmark \\ 
 
 Citrus \cite{Citrus}  & $\sim$ & $\times$ & \checkmark \\ 
 
 SoapUI \cite{soapUI}  & $\sim$ & $\times$ & \checkmark \\ 
 
  Wilma \cite{Wilma} &  $\sim$ & $\times$ & \checkmark\\ 

 Whole-Cluster OSV \cite{du:2013} &  \checkmark & $\times$ & $\times$ \\ 

Entropy-Weighted OSV  \cite{versteeg2016enhanced} & \checkmark & $\times$ & $\times$ \\

 Cluster-Centroid OSV \cite{du:2015interaction}  & \checkmark & $\times$ & $\times$ \\

Prototype OSV \cite{versteeg:2016} & \checkmark & $\times$ & $\times$ \\

 SSBV \cite{enicser2018testing} & \checkmark & $\sim$ & $\times$ \\
 \hline

This paper & \checkmark  & \checkmark & \checkmark \\

\hline
 
\end{tabular}
}
\end{center}
\end{table}

\subsection{Service Virtualization Techniques}
\label{subsec:servicevirtualisation}
Service virtualization involves the creation and deployment of \dquotes{virtual services} that emulate the specific behavior of the actual connected services and facilitate the testing of the system under test without requiring access to the actual services. There have been a number of efforts in creating service virtualization techniques and tools for software testing (see Table~\ref{tab:relatedwork}). Several commercial service virtualization tools are available, including CA Service Virtualization \cite{caservicevirtualization}, ServiceV Pro \cite{serviceVPro}, and IBM Service Virtualization \cite{IBMservicevirtualization}. There are also open-source tools such as WireMock \cite{wiremock}, Hoverfly \cite{hoverfly}, Citrus \cite{Citrus}, SoapUI \cite{soapUI}, and Wilma \cite{Wilma}. However, much manual work is required to define or infer the behavior of the service being virtualized or modeled. In addition, none of these approaches can achieve the necessary level of accuracy in emulating the service behavior, especially for stateful services. 

Research efforts have explored the automated derivation of service models from service interaction traces for use in service virtualization \cite{du:2013, du:2015interaction, versteeg2016enhanced, versteeg:2016}. These approaches leverage algorithms from bioinformatics to automatically analyze service interactions and derive service models. This is particularly valuable because manually defining service models \cite{hine:2016} is often labor-intensive \cite{versteeg:2016} and might not be practical due to insufficient information available, especially in the case of legacy systems \cite{ghosh:1999}.

The \textit{Whole-Cluster} Opaque Service Virtualization (OSV) approach \cite{du:2013} employs the Needleman-Wunsch algorithm \cite{needleman:1970general} to identify a service request in the interaction trace that is most similar to the incoming request and generates a response by substituting the payload in the corresponding response of that identified request. However, this approach often produces inaccurate responses, especially when different types of request messages contain similar payloads. 
The \textit{Entropy-Weighted} OSV approach \cite{versteeg2016enhanced} addresses this issue by assigning more significance to the operation/request type field over other message fields, improving the accuracy in finding the closest match for the incoming request. Despite this improvement, it remains inefficient in generating responses, as it takes considerable time to locate the closest matching request, especially with large interaction traces. 

The \textit{Cluster-Centroid} OSV approach \cite{du:2015interaction} tackles this efficiency issue by clustering requests based on a dissimilarity ratio matrix and selecting a centroid request for each cluster. An incoming request is  compared to these centroid requests rather than every request in the recorded trace, significantly enhancing runtime performance. However, it often does not match the incoming request to the most similar requests in the trace, 
resulting in less accurate responses than the Whole-Cluster OSV approach.
The \textit{Consensus} or \textit{Prototype} OSV approach \cite{versteeg:2016} offers further improvements by creating a consensus message prototype for each cluster (i.e., for each type of request message) and comparing incoming requests to these prototypes. This approach enhances accuracy compared to the Cluster-Centroid OSV approach. 
In general, all these OSV approaches generate responses based solely on the incoming request without considering the service's state or state changes over time. 
Consequently, they are unable to synthesize responses for stateful services accurately.

\begin{figure*}
	\centering
	\includegraphics[width=.9\textwidth]{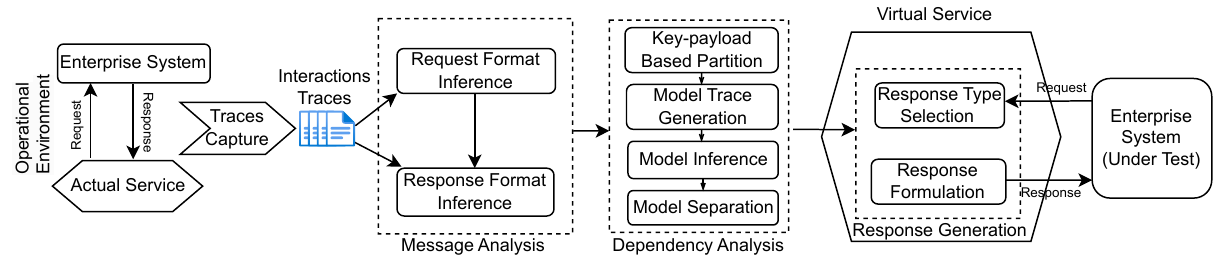}
	\caption{Approach overview}
	\label{fig:overview}
\end{figure*}

To our knowledge, the only techniques that account for service state when synthesizing responses are presented in \cite{enicser2018testing}. It utilizes classification-based virtualization (CBV) and sequence-to-sequence-based virtualization (SSBV) to virtualize stateful services. In CBV, response messages are labeled based on sequences of k preceding request messages, and a decision tree classifier is trained to map these request sequences to corresponding response types. At runtime, each incoming request is encoded with the previous k interactions (both requests and responses) and compared to the training data to predict the response. 

SSBV uses long short-term memory (LSTM), a neural network technique, to model services. Input sequences (messages) are transformed into state vectors, and an LSTM-based sequence-to-sequence (Seq2Seq) model is trained to predict the next character in the target sequence based on the preceding characters. During response generation, a single-character sequence is encoded into a state vector, which is fed into the decoder to predict the next character. This character is appended to the sequence, and the process is repeated until the entire sequence is generated. However, because responses are predicted character by character, the accuracy diminishes when the response type is more than one character long. 
Both techniques do not consider or separate the service states tied to specific data records when generating responses. As service states can differ for different data records, as explained in Section~\ref{sec:motivation}, the effectiveness of these methods is limited when virtualizing stateful services.

In contrast to the above approaches, our goal is to emulate the stateful services by deriving a more accurate service model that can keep track of the service state and its changes, by capturing the inter-message dependency and the message-data dependency across service interactions. First, the derived service model needs to identify the sequence of interactions responsible for changing the state of the service and consequently consider the service state in identifying the response message type. Furthermore, the derived service model also needs to consider the message payloads and their relationships across messages as the synthesized response should be about the appropriate service data record(s) and contain relevant payloads, reflecting another aspect of the service state. Finally, the message formats need to be considered in formulating response messages to increase their accuracy.

\subsection{Techniques for Inferring System Control Models}
\label{subsec:modeinference}
Part of a service or system's behavior model is its control dependency, \ie, what \textit{types} of messages or message sequences are allowed at a given time, which usually reflects and depends on the service's state. There have been many existing efforts in inferring the control dependency model from a system's log or interaction trace. Some of the works consider only the normal system interactions as input, while others instrument the program code to incorporate additional information in the system log.


The kTail algorithm \cite{biermann:1972} infers a system model from a system log by merging equivalent states. The technique proposed in \cite{Wang:2013} further considers the `immediate follow' invariants in merging the states, which produces a more generalized model compared to the model generated by kTail. The technique proposed in \cite{krueger:2012} considers future temporal rules to prevent imprecise state merging and consequently, infers a more precise system model compared to kTail. The technique proposed in \cite{lo:2009} mines both future and past temporal rules from the input trace and uses them in the state merging step of kTail to produce a more precise system model. 
Synoptic \cite{beschastnikh2011b}, Perfume \cite{ohmann:2014}, and CSight \cite{beschastnikh:2014} infer a system model from a system log by first creating a separate path for each input sequence, then forming a coarse-grained initial model, and finally using the CEGAR \cite{clarke:2000} technique to refine the model utilizing temporal invariants mined from the input log.  

Several other research efforts have also considered data values in deriving the control model of a system from its log. They combine data values and method invocation sequences in the model inference process. The techniques proposed in gkTail~\cite{lorenzoli:2008} and \cite{walkinshaw:2016} infer a finer-grained system model in the form of Extended Finite State Automata (EFSA) by considering data values' impact on the next event (message). An empirical study comparing the techniques that infer a simple FSA model and the techniques that infer an extended FSA model with information about data values shows that considering data values together with invocations can represent the behavior of the actual system more accurately \cite{lo:2012}.

These techniques are useful in determining the dependencies between messages of a software service. However, none of them is aimed at generating responses for incoming requests. Furthermore, some of them only consider the method invocation sequences in inferring the system model and ignore the complex interplay between the data values and the method invocations. 
In the service context, however, the data values of messages play an important role in determining the type and payload of the response for an incoming request.

The existing techniques that consider the effect of data values on the method invocation sequence require the instrumentation of the program code of the targeted system or service. Such access to the program code limits their applicability in inferring control dependency models where access to the program code for instrumentation is not possible. As such, these techniques are not directly applicable in service virtualization as only interactions between a client and the actual service are accessible and the program code implementing the targeted service is not available.



\section{Approach}
\label{sec:approach}
Our approach to service virtualization starts with taking an interaction trace of a service as input, proceeds to derive a behavior model for the service, and ultimately uses this model to generate responses to incoming requests. This derived service model can facilitate the testing of an enterprise system (the SUT) without needing access to the connected service. 
As illustrated in Fig.~\ref{fig:overview}, our approach has three main stages: message analysis, dependency analysis, and response generation. The message analysis and dependency analysis stages are carried out offline, while the response generation stage occurs at runtime when the service model is used for testing the SUT.

\subsection{Message Analysis - An Overview}
\label{subsec:traceAnalysis}
To derive a behavior model for a service from its interaction trace, our approach first takes the interaction trace as input, identifies the types of request and response messages within it, and infers their formats. 
This subsection provides an overview of how the message types and message formats are discovered, while more detailed explanations can be found in~\cite{hossain2022extracting} and~\cite{jiang:2019positional}.

A service usually supports different types of request and response messages, and therefore the interaction trace collected from the service contains request and response messages of these different types. To distinguish and formulate the service's messages, the types and formats of the messages in the trace need to be identified. By following the technique proposed in~\cite{hossain2022extracting}, 
the message fields for the request message types can be identified by aligning the stable message fields across all request messages and analyzing the fields' occurrence rates. Then, the request message type fields are used to classify the request messages into type-specific clusters. Finally, the format for each request message type or cluster is derived by identifying the recurring keyword pattern across the messages in each cluster. For example, this technique can identify five request message types, \messagetype{Bind, Add, Delete, Search} and \messagetype{Unbind}, from the request messages of Table~\ref{tab:exampleinteractionlibrary}. The message formats for each of the request message types are further inferred. For example, the \messagetype{Search} request messages (\ie, the messages with {\tf id: 23, 130, 135, 144 and 251}) has the following format:

\begin{center}
\tracefont{\{id:(.*),op:S,cn:(.*)\}}
\end{center}

A service not only accepts different types of request messages but also sends different types of response messages for the request messages. Furthermore, the same type of requests may lead to different types of responses, depending on the service's state. 
As the response messages often do not have a distinctive type field, 
a different message clustering technique is required to identify the response message types and their formats. 
In this regard, the P-token technique~\cite{jiang:2019positional} is adopted:
it classifies the response messages into type-specific clusters by identifying the co-occurrence patterns of common message fields (\ie, the response message types) across the response messages of those request messages of a given type; then, the format for each type or cluster of response messages is inferred by identifying the common keywords over the messages in the cluster. For example, the P-token technique can identify two response types, \ie, \messagetype{Ok}, and \messagetype{Not Found} from the response messages of the \messagetype{Search} request type. The following message format is inferred for the \messagetype{Ok} response messages (\ie, the messages with {\tf id: 130 and 251}). 
\begin{center}
\tracefont{\{id:(.*),op:SearchRsp,result:Ok,cn:(.*), sn:(.*),mobile:(.*)\}}
\end{center}

 After the types and formats of the request and response messages are identified, each interaction in the interaction trace is labeled with its \emph{type} (composed of the request type and response type). The interaction type, along with the inferred format of its response messages, is stored as a key-value pair in the \emph{ResponseMap}, which is used for response generation at runtime. For example, the above response format is stored in the ResponseMap with the key  \messagetype{S\_SearchRsp(Ok)},  where \messagetype{S\_ SearchRsp(Ok)} is the interaction type for the response cluster that contains the interactions with {\tf id: 130 and 251}.

\subsection{Dependency Analysis}
\label{subsec:serviceModelInference}
The second stage of our approach focuses on inferring the service's behavior model from its interaction trace, which will be used to generate response messages at runtime. The service model takes into account the dependencies between different message types in relation to the application data records, capturing the service's interaction behavior and its reliance on both the service state and state changes.

The dependency analysis process comprises four key steps:
(1) the interaction messages in the trace are segmented based on their key payloads, partitioning the interactions into groups according to the the data entities or records they are concerned; 
(2) the interactions within each group are replaced with their interaction types, creating a simplified type-based model trace;  
(3) the overall message dependency model of the service is derived from the model trace, capturing the relationships and dependencies among the service messages; 
(4) the inferred message dependency model is divided into two sub-models: one representing messages that include key payloads and another for messages without key payloads.
The specifics of these steps are captured in Algorithm~\ref{alg:messageDependencyModelInference} and further discussed in the following subsections. 

\subsubsection{Service Trace Partition}
\label{subsec:recordBasedPartitionChapter6}
As observed in Section~\ref{sec:motivation}, a service usually manages a collection of data records, where each record follows its own distinct life cycle, independent of other records. To obtain a more precise understanding of the dependencies between messages, the service trace is first divided into sub-traces specific to each data record, based on their key payloads.

Based on the message formats identified in the previous stage, a user-defined regular expression is formulated to identify the key payload in the request messages, and used to partition the interaction trace into sub-traces with each containing the interactions under a specific key payload (see lines \emph{4} to \emph{13} in Algorithm~\ref{alg:messageDependencyModelInference}). First, the regular expression is used to parse all the interactions (or more precisely their request messages) and extract the key payloads associated with the request messages (line \emph{5}). For example, \messagetype{Judith} is extracted as the key payload from the second interaction (\tf{id:2}) of Table \ref{tab:exampleinteractionlibrary} for the given regular expression \messagetype{cn=(.*?)}. The interaction is then placed into a hashmap based on its key payload (line \emph{11}). 

Certain interactions do not have a key payload and do not concern a specific data record, as they serve purposes like authentication and are not associated with any specific data record. For instance, the \messagetype{Bind} and \messagetype{Unbind} requests in Table \ref{tab:exampleinteractionlibrary} are used to connect and disconnect a user to or from the directory server. 
On the other hand,  
(\messagetype{Bind}) has a contextual dependency with other requests, meaning the directory service only permits the insertion or modification of entries after successful user authentication via a \messagetype{Bind} request. 
To capture this dependency, interactions without a key payload are replicated across each data record-specific sub-traces (lines \emph{6} to \emph{9}).

Table \ref{tab:recordbasedpartition} shows the result of partitioning the interactions in Table \ref{tab:exampleinteractionlibrary} based on the key payloads. Partitions 1, 2, 3, and 4 are generated for the key payloads \messagetype{Judith}, \messagetype{Gavin}, \messagetype{Linden} and \messagetype{Katy}, respectively. Note that interactions with \tf{id: 1} (Bind), \tf{id: 96} (Unbind), \tf{id: 105} (Bind) and \tf{id: 300} (Unbind) appear in every sub-trace.   

\begin{table}[!t]
    \centering
    \caption{Partitioned interaction trace}
    \label{tab:recordbasedpartition}
    \includegraphics[width=.4\textwidth]{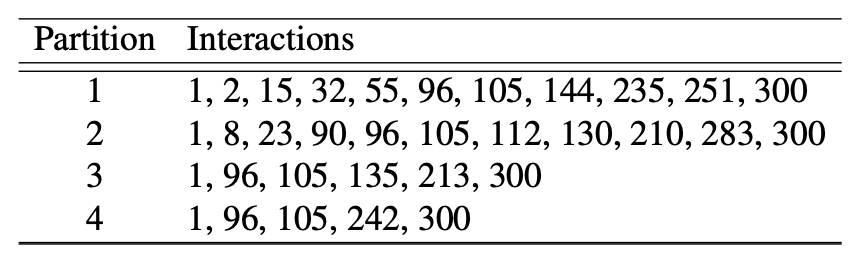}
\end{table}

\subsubsection{Model Trace Generation}
The message dependency model concerns the sequential relationships between the types of the messages or interactions in the service trace, abstracting away the specifics of the message payload values (including key payloads). This step converts and merges the message trace partitions from the last step into a model trace 
in terms of interaction types. 

First, a model trace of interaction types for each key-payload based partition is formed, by replacing each interaction in the petition with its type,
and inserting a separator (,) between them (see lines \emph{14} to \emph{20} of Algorithm~\ref{alg:messageDependencyModelInference}). For example, the model trace in the first row of Table \ref{tab:eventTraces} is created from the interactions of partition 1 of Table \ref{tab:recordbasedpartition}; similarly, the model traces in the second, third, and fourth rows are created from the interactions of partitions 2, 3, and 4 of Table \ref{tab:recordbasedpartition}, respectively. 

Finally, the overall model trace is formed as a trace of interaction types, by sequentially putting together all the model traces for the different partitions (see lines \emph{14} to~\emph{20}). 
Table \ref{tab:eventTraces} presents the overall model trace generated from the interactions of Table~\ref{tab:exampleinteractionlibrary}, after all the key-payload based model traces are sequentially concatenated. Unlike the actual interaction trace, the model trace omits payloads and other message-specific details, preserving only the request and response types. As a result, the model trace captures the relationships between the service interactions in terms of their types, but executed sequentially on individual data records in turn without any interleaving.

\begin{table*}[!t]
    \centering
    \caption{Model interaction trace}
    \label{tab:eventTraces}
    \includegraphics[width=.9\textwidth]{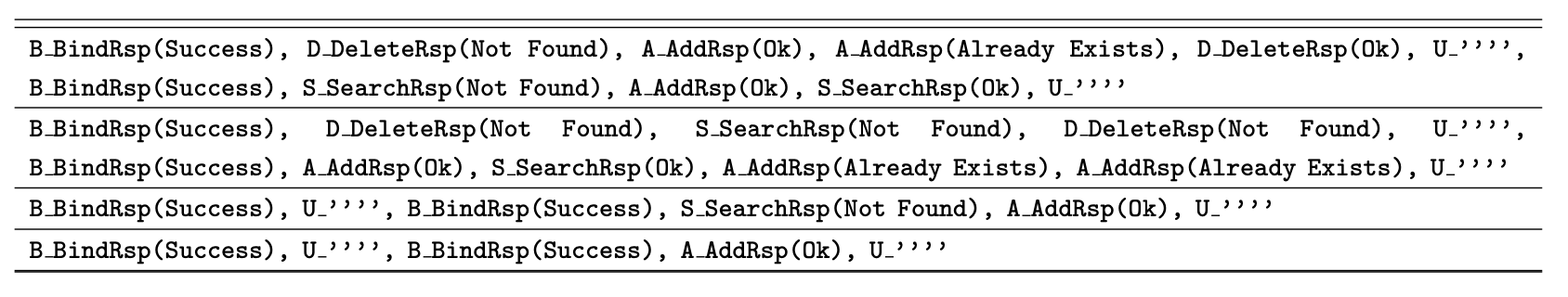}
\end{table*}

{\setlength{\intextsep}{0pt}
\begin{algorithm}[!t]
	\caption{Message dependency model inference}\label{alg:messageDependencyModelInference}
\begin{algorithmic}[1]
	\STATE \textit{Input:} Interaction trace \emph{T}, regular expression \emph{r}
	\STATE \textit{Output:} a set of message dependency model(s): \emph{M(FSA, FSA1, FSA2)}
	\STATE \textit{Initialization:}  Key payload set \emph{K} =$getAllKeyPayloads(T)$, partition hashset \emph{H(K)}= \emph{$\emptyset$}, model  trace \emph{MT} =$emptyString$
	\FOR{$Interaction$ $I \in T$} $  $ 
	     \STATE $k$ $\leftarrow$ $extractKeyPayload(I,r)$  $  $ 
               \IF{$length(k) = 0$} $  $ 
	               \FOR{$k2 \in K$} 
                             \STATE $H(k2) \leftarrow H(k2)$ $U$ $\{I\}$
                \ENDFOR
                \ELSE  $  $ 
	               \STATE $H(k) \leftarrow H(k)$ $U$ $\{I\}$
	    \ENDIF
	\ENDFOR
	\FOR{$k \in K$} $  $
	    \FOR{$I \in H(k)$}
	        \STATE $IT \leftarrow getInteractionType(I)$ $  $
	        \STATE $MT \leftarrow$ $StringConcat(MT,IT, ',')$ $  $
	    \ENDFOR
	    \STATE $MT \leftarrow$ $ReplaceLastChar(MT,newline, ',')$ $  $
	\ENDFOR
	\STATE $FSA$ $\leftarrow$ $generatePTA(MT)$ $ $
	\FOR{$s1 \in states(FSA)$ $AND$ $s1 \neq initial$} $ $
	        \FOR{$s2 \in states(FSA)$ $AND$  $s1 = s2$ }
	                \STATE $FSA \leftarrow$ $Merge(FSA, s1,s2)$ 
	        \ENDFOR
	\ENDFOR
    \STATE $FSA1 \leftarrow FSA$; $FSA2 \leftarrow FSA$ $ 
 $
	\FOR{$s \in states(FSA)$}
	    \IF{$IsKeyPayload(s)$}
	        \STATE $FSA2 \leftarrow Remove(FSA2, s)$  $ 
 $
	    \ELSE   
	        \STATE $FSA1 \leftarrow Remove(FSA1, s)$ $ 
 $
	    \ENDIF
	\ENDFOR
    \STATE $M(Full, Key,nonKey) \leftarrow$ $(FSA,FSA1,FSA2)$ $ 
 $
	\STATE $Return$ $M$ $ 
 $
\end{algorithmic}
\end{algorithm}
}

\subsubsection{Dependency Model Inference}
\label{subsec:modelInference}
\begin{figure*}[!t]
	\centering
	\includegraphics[width=1\textwidth]{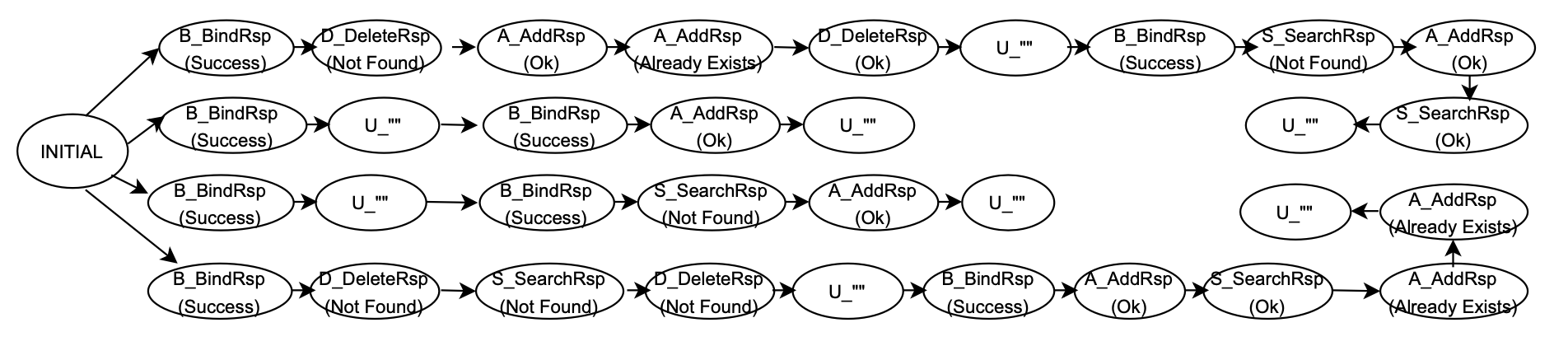}
	\caption{Prefix tree acceptor (PTA)}
	\label{fig:pta}
\end{figure*}

The objective of this step is to infer the service's message dependency model from the model trace obtained in the last step. The inferred model captures the dependency between messages (or more precisely, between message types) and consequently, is used to track the current state of the services at run-time. 
First, a prefix tree acceptor (PTA) is generated from the model trace, where a separate path is created for each interaction sequence of the model trace (line \emph{21}). For example, Fig.~\ref{fig:pta} shows the PTA generated from the model trace in Table~\ref{tab:eventTraces}. Then, the kTail algorithm with $k=0$ \cite{biermann:1972} is used to merge the equivalent states in the PTA and infer the message dependency model (lines \emph{22} to \emph{26}). All of the outgoing edges of the merging states are carried over into the merged model, and therefore, the merged model accepts all of the traces that are accepted by the PTA.

The kTail algorithm can be used with different $k$ values. This means that $k$ preceding and subsequent events (interactions) will be compared in merging states, reflecting the sequential/temporal constraints on method invocations (request messages). 
In a service environment, a client can send any request to a service at any time (i.e., without adhering to a specific order) and will always receive a response back from the service, including the possibility of receiving an error response if the client issues an invalid request at that moment. That is, there are constraints on the order of the request messages, but only with the possibility of having different types of responses.
As such, the kTail algorithm with $k=0$ is used in our approach to infer a service's message dependency model.

Fig.~\ref{fig:inferredModel} shows the inferred message dependency model after merging equivalent states of the PTA in Fig.~\ref{fig:pta}. It is important to note that the example dataset contains a very limited number of interactions and does not cover all types of possible interaction sequences permitted by the service. As a result, the inferred model in Fig.~\ref{fig:inferredModel} does not include all potential interaction paths. In a more realistic scenario, however, the inferred model would feature more states and edges, often resulting in a configuration where every state is connected to every other state (i.e., a flower model), including interactions that involve error responses. For example, the inferred message dependency model for the full CA IM (LDAP) dataset (see Section \ref{subsec:datasets}), generated using our approach, comprises \emph{15} states and \emph{117} edges.

\subsubsection{Dependency Model Separation}
\label{subsubsec:splitControlModel}

The overall message dependency model derived from a service's interaction trace in the previous step represents the valid sequences of interactions or messages \textit{in a session} of using the service. For example, the dependency model for the directory service captures the relationships between authentication requests (such as \messagetype{Bind} \messagetype{unbind}) and other requests for normal directory operation (such as \messagetype{Add} \messagetype{Modify}). 
On the other hand, the interactions without key payloads (\eg, \messagetype{Bind} and \messagetype{Unbind}) do not concern the data records of the service (part of the service state) while others do. For the purpose of capturing the service data record changes and their impact on service interactions \textit{across multiple sessions}, the overall message dependency model is split into two sub-models: one for the interactions with key payloads and the other for the interactions with no key payloads. This facilitates the use of the message dependency model in response generation (see below). 

\begin{figure}[!t]
\centering
\begin{subfigure}{0.42\columnwidth}
\includegraphics[width=.9\textwidth]{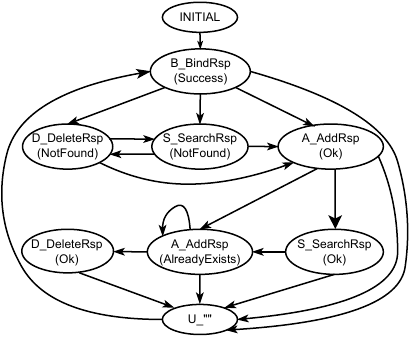}
    \caption{Overall message dependency model}
    \label{fig:inferredModel}
\end{subfigure}
\hfill
\begin{subfigure}{0.4\columnwidth}
\includegraphics[width=.96\textwidth]{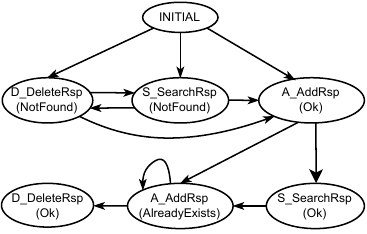}
    \caption{Key payload}
    \label{fig:inferredModelKeyPayload}
\end{subfigure} 
\hfill
\begin{subfigure}{0.15\columnwidth}
\includegraphics[width=.88\textwidth]{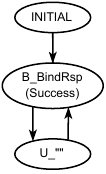}
    \caption{Non-key payload} 
    \label{fig:inferredModelLDAPBindUnbind}
\end{subfigure} 
 \caption{Inferred message dependency models}
 \label{fig:inferredModels}
\end{figure}

In splitting the message dependency model, the sub-model for messages with data records is formed by removing the states and their edges that do not concern a data record from the inferred overall model, and form the separate sub-model for messages not concerning data records by combining those removed states (see lines \emph{27} to \emph{34} of Algorithm~\ref{alg:messageDependencyModelInference}).
For example, the message dependency model in Fig.~\ref{fig:inferredModel} is split into a sub-model (key payload) for messages concerning data records in Fig.~\ref{fig:inferredModelKeyPayload} and a sub-model for messages not concerning data records in Fig.~\ref{fig:inferredModelLDAPBindUnbind}. 
In particular, the latter contains only those states and edges that are related to \messagetype{Bind} and \messagetype{Unbind} and are detached from the full message dependency model when extracting the key payload-related sub-model.

\smallskip

\subsection{Response Generation}
\label{subsubsec:responseGeneration}
Once synthesized from a service's interaction trace, the service's dependency model can be used in a corresponding virtual service to generate responses at runtime for incoming requests from the system under test (SUT). This response generation method has two steps: (1) determining the appropriate type of response message for the incoming request, and (2) populating the response message with suitable payload data.

\subsubsection{Response Type Selection}
\label{subsec:responseTypeSelection}
 
For service state-based response generation at runtime, the service state is maintained and tracked as the service interactions progress, especially recognizing that different data records of the service may be in different states at a given time. Between the two inferred message dependency sub-models (see Fig. \ref{fig:inferredModels} for example), the key payload-based sub-model is applied for requests that contain a key payload, and the non-key payload sub-model is used for requests that do not have a key payload (see Algorithm~\ref{alg:responseSelection}). In particular, the appropriate response type for an incoming request is determined by evaluating the service state specific to the data record concerned, using the payload-based message dependency sub-model.

It is important to note that a payload-based service dependency sub-model instance is created and tracked for each data record (or key payload) as relevant requests from the SUT appear during interactions with the virtual service. One sub-model instance is also created and maintained for non-key payload interactions as such requests occur. More specifically, the current state of each key payload-based service sub-model instance is stored and managed in a key-value state map \emph{H} (see line \emph{1} of Algorithm \ref{alg:responseSelection}) whenever a request is received from the SUT.

The request type and key payload are identified from the incoming request at lines \emph{4} and \emph{5}, respectively.
At line \emph{6}, the algorithm checks if a model instance associated with the current request (whether with a key payload or not) is already created and stored in the state map (\ie, through a previous request).
If such an entry is found in the map, the current state of the service (\emph{c}) is set with the stored state (line \emph{7}); otherwise, the `INITIAL' state is set as the current state for the respective new model instance (lines \emph{9} to \emph{13}). 
 
At line \emph{15}, the interaction type (\emph{I}) is initially set to an empty string for the request message (\emph{rq}). The next step involves examining the outgoing edges of the current node in the message dependency sub-model to locate a connected node with a label that matches the incoming request type (lines \emph{16} to \emph{23}). If a matching node is found, the state of the related sub-model instance in the state map is updated to reflect this new state, and the interaction type for the request (\emph{I}) is set to the label of the selected node (lines \emph{19} to \emph{21}). Lastly, at line \emph{24}, the interaction type (\emph{I}) (importantly, including the response type) for the incoming request is returned.

An empty string will only be returned if the dependency model lacks a matching path for the incoming request, meaning that the request message does not appear in any invocation sequence within the service trace. For example, a \messagetype{delete} request may follow an \messagetype{add} request at runtime; however, if the interaction trace used to build the model has never showed a \messagetype{delete} request following an \messagetype{add} request, the model would not recognize this sequence. Such cases are uncommon when the service model is based on a comprehensive interaction trace, and in fact, they did not occur in our experiments.

\begin{algorithm}[!t]
	\caption{Response type selection}\label{alg:responseSelection}
\begin{algorithmic}[1]
	\STATE \textit{Input:}  request message \emph{rq}, message dependency models \emph{M(Full, Key, NonKey)}, key payload and service state map \emph{H}
	\STATE \textit{Output:} interaction type \emph{I} (for the request message)
	\STATE \textit{Initialization:} node labels (interaction types) \emph{L}=getNodeLabel(G) 
	\STATE $r$ $\leftarrow$ $getRequestType(rq)$ 
	\STATE $d$ $\leftarrow$ $getKeyPayload(rq)$ 
        \IF{$d \in H.keys $} 
	   \STATE $c \leftarrow H.get(d)$ $ $
        \ELSE  $  $ 
        \IF{$length(d) > 0 $}  $  $ 
	    \STATE $c \leftarrow M(key).INITIAL$
	\ELSE
	   \STATE $c \leftarrow M(nonKey).INITIAL$
	\ENDIF
	 \ENDIF

      \STATE $I \leftarrow $ 	$emptyString$ $   $ 
	 
	 \FOR{$e \in c.outEdges$} $  $ 
	    \STATE $n \leftarrow e.Target$  $  $ 
	    \STATE $l \leftarrow L.get(n)$ $  $ 
	    \IF{$r \in l$}  $  $ 
	        \STATE $H$ $\leftarrow$ $push(d,n)$  $ $  
	         \STATE $I \leftarrow $  $l$ $ $ 
	    \ENDIF
	 \ENDFOR
	\STATE $Return$ $I$ $ $ 
\end{algorithmic}
\end{algorithm}

As an example, let us consider the following \messagetype{Delete} request (second request in Table \ref{tab:exampleinteractionlibrary}): 


\centerline{\tracefont{\{id:2,op:D,cn:Judith\}}}

\vskip 0.1in

\noindent According to Algorithm \ref{alg:responseSelection}, request type \dquotes{\messagetype{D}} and \dquotes{\messagetype{Judith}} are extracted as the request type and key payload respectively. It then attempts to locate the current state of the payload \dquotes{\messagetype{Judith}} in the state map (or corresponding model instance). Since this is the first time the key payload \dquotes{\messagetype{Judith}} appears, there is no existing model instance, and as a result, the current state (\textit{c}) is set to ``\messagetype{INITIAL}'', signifying the creation of a new model instance for \dquotes{\messagetype{Judith}}.

Next, the algorithm checks all outgoing edges from the initial state to find a connected node that matches the incoming request type, \dquotes{\messagetype{D}} (\ie, it identifies the connected node ``\messagetype{D\_DeleteRsp (NOT Found)}", which contains the request type \dquotes{\messagetype{D}}). The current state of the newly created model instance for \dquotes{\messagetype{Judith}} is then updated to this node, \dquotes{\messagetype{D\_DeleteRsp (NOT Found)}}, and this state is added to the service state map \emph{H}. Finally, ``\messagetype{D\_DeleteRsp (NOT Found)}" is returned as the interaction type for this incoming request.

\subsubsection{Payload Population}
\label{subsubsec:responseTransformation}
After determining the response time (from the interaction type) in the previous step,  this step populates the response message with the appropriate payloads (see Algorithm \ref{alg:populateResponseMessage}). It takes as input, the request message, the interaction type, the interaction trace, the response message formats for each request type (ResponseMap), and the payload equality rules (PayloadEqualityMap) (see below). It then generates  a response message after inserting the appropriate payloads into the response type identified. 

Line 4 of Algorithm \ref{alg:populateResponseMessage} extracts the request type from the incoming request. As noted previously, an empty string may be returned as the interaction type if the message dependency model doesn’t recognize the request message at its current state. In such a case, the interaction type is randomly selected from all possible interaction types that correspond to the incoming request type (lines 5 to 7), to return some possible response (even though not identified by the dependency model or interaction trace) rather than nothing. For example, if Algorithm \ref{alg:populateResponseMessage} returns an \emph{empty} string for a \messagetype{Delete} request, 
\messagetype{D\_DeleteRsp(NotFound)} or  \messagetype{D\_DeleteRsp(Ok)} will be randomly selected.

\begin{algorithm}[!t]
	\caption{Payload population}\label{alg:populateResponseMessage}
\begin{algorithmic}[1]
	\STATE \textit{Input:} request message \emph{rq}, interaction type \emph{I}, interaction trace \emph{T}, ResponseMap \emph{R}, PayloadEqualityMap \emph{S} 
	\STATE \textit{Output:} response message $response$
	\STATE \textit{Initialization:} message field set \emph{field} = \emph{$\emptyset$}, payload set \emph{payload} = \emph{$\emptyset$}, chosen response message \emph{y} = ""
    \STATE $r$ $\leftarrow$ $getRequestType(rq)$
    \IF {$IsEmpty(I)$}
        \STATE $I$ $\leftarrow$ $selectInteractionTypeRandomly(R, r)$
    \ENDIF
	\STATE $p$ $\leftarrow$ $getResponseType(I)$ 
	\STATE $f$ $\leftarrow$ $R.get(I)$  
    \STATE $ field \leftarrow $  split$(f, (.*)) $
    \STATE $Y$ $\leftarrow$ $selectAResponseRamdomlyOfGivenType(T,I)$
    \STATE $q$ $\leftarrow$ $S.get(I)$  
    
	\FOR{$i \in \{1,\dots,|field|\}$}
	    \IF {$IsContained(i, q)$}  
	        \STATE $payload(i)$ $\leftarrow$ $extractPayload(rq,i)$
            \ELSE
                \STATE $payload(i)$ $\leftarrow$ $extractPayload(y,i)$
	    \ENDIF
	\ENDFOR
	
	\STATE $response$ $\leftarrow$ $insertPayload(payload,f)$
	\STATE $Return$ $response$
\end{algorithmic}
\end{algorithm}

Next, the response type is extracted from the interaction type (line 8), and the response message format is accessed from the \emph{ResponseMap} using the selected interaction type (line 9). For instance, if \dquotes{\messagetype{D\_DeleteRsp(Not Found)}} was chosen for an incoming \messagetype{Delete} request (line 6), the format of the response message would be \dquotes{\tracefont{{id:(.*),op,result Found}}}.

Once the response format is identified, the variable portions meant for holding payloads are extracted (line \emph{10}). For each message field in the chosen response type, the payload is filled using either (1) the incoming request message or (2) a randomly selected response message from the interaction trace that corresponds to the identified interaction type.


\textit{(i) Payload from incoming request:}
The payload substitution technique from~\cite{du:2013} is adopted to 
identify the payload equality relationships across the request and response fields of the same interaction. These relationships are stored as equality-based substitution rules in a \emph{PayloadEqualityMap} (line \emph{1}). Based on these equality relationships (lines \emph{12} to \emph{14}), the relevant payloads from the incoming request are extracted, and used in the respective response fields (line \emph{15}) according to the response format. For the example incoming request above, using the equality-based substitution rule between the message IDs of the request and response messages, the message ID ``\tracefont{2}'' is extracted from the example incoming request and is used as the value or payload (\ie ``\tracefont{2}'') for the message ID field of the generated  response message, resulting in a message \textit{identical} to the expected response: 

 \begin{center}
     \tracefont{\{id:2,op:DeleteRsp,result:Not Found\}}
 \end{center}

The above-generated response does not contain any other data fields that are not in the request messages, and hence, there is no need to use the interaction trace to determine any other payloads in the above response.

\textit{(ii) Payload from randomly selected interaction:}
When the response message contains data fields that are not filled according to the payload substitution rules,  one of the responses of the identified interaction type (line \emph{11}) is randomly selected, and its relevant field values from that response are extracted as approximations to the expected payload values in the generated response message (line \emph{17}). The selection is done radomly because our experiments show that the closest matched interaction as described in \cite{du:2015interaction} does not improve the accuracy but reduces the efficiency. 

As an example, let us consider the incoming request and the expected or actual response below:

\setlength{\tabcolsep}{0.2em}
\begin{table}[!ht]
    \centering
    \begin{tabular}{lm{8.2cm}}
        Incoming Request: & \tracefont{\{id:345,op:S,cn:Craig\}}  \\ 
        Expected Response: & \tracefont{\{id:345,op:SearchRsp,result:Ok, sn:LINK,mobile:543219087\}}
    \end{tabular}
\end{table}
\setlength{\tabcolsep}{0.45em} 
\FloatBarrier

\noindent The response type \messagetype{S\_SearchRsp(Ok)} is selected according to the message dependency model (line \emph{9} of Algorithm \ref{alg:populateResponseMessage}). 
It has the following response format (line \emph{9}): 

\setlength{\tabcolsep}{0.2em}
\begin{table}[h]
    \centering
    \begin{tabular}{m{1.5cm}m{0.1cm}m{6.9cm}}
       Response Format: & & \tracefont{\{id:(.*),op:SearchRsp,result:Ok,cn:(.*), sn:(.*),mobile:(.*)\}}  \\
    \end{tabular}
\end{table}
\setlength{\tabcolsep}{0.45em} 

\noindent \dquotes{\tracefont{345}} and \dquotes{\tracefont{Craig}} are extracted from the incoming request using the substitution technique~\cite{du:2013}, and are used as the values for the first and second payload fields in the generated response. 
As the remaining two fields of the response message do not have corresponding payloads from the incoming message, an interaction is randomly selected from all the \messagetype{S\_SearchRsp(Ok)} messages in the interaction trace (say, interaction with {\tf id: 130} is selected from the interactions with {\tf ids:130, 251}). 
 The values of its third and fourth payload fields (\ie, \dquotes{\tracefont{MAJOR}} and \dquotes{\tracefont{26952135}}) are extracted, and inserted into the corresponding positions in the generated response. The generated response will have the following formulation: 
 
  \begin{center}
    \tracefont{\{id:345,op:SearchRsp,result:Ok,cn:Craig, sn:MAJOR,mobile:26952135\}}
  \end{center}    

The synthesized response above includes the expected payloads in the \messagetype{id} and \messagetype{cn} fields, but not in the \messagetype{sn} and \messagetype{mobile} fields. Therefore, it is considered a close approximation to the expected response, i.e., data-consistent according to the evaluation criteria outlined in Section \ref{subsec:evaluationmetric}.

\section{Probabilistic Message Dependency Model}
\label{sec:probabilisticModel}

The inferred message dependency model shown in Fig.~\ref{fig:inferredModel} is deterministic because the traces in Table \ref{tab:exampleinteractionlibrary} were collected under the assumption that the initial state of the service data records is empty (hereafter referred to as \textit{Clean Start}). In reality, the service state is often \textit{non-empty} (hereafter referred to as \textit{non-clean Start}) at the beginning of data collection. The data records involved in the captured interaction trace may or may not have been added before the data collection commenced. These records are crucial in determining the responses to incoming requests, particularly for stateful services.

For instance, if the record \dquotes{\messagetype{Judith}} already existed before the trace collection, the response for the \messagetype{Delete} request (id:2) in Table~\ref{tab:exampleinteractionlibrary} would be \messagetype{Ok} rather than \messagetype{Not Found}. In this scenario, the subsequent responses for requests with the key payload \dquotes{\messagetype{Judith}} would also differ from those presented in Table~\ref{tab:exampleinteractionlibrary}. This indicates that the same request can yield two or more alternative responses based on the service's state at the time of trace collection. To address these variations, a probabilistic control model is derived from the interaction traces, \ie, attaching probabilities of the relevant response types in the message trace to the transitions within the inferred model.

\begin{figure}[!t]
	\centering
	\includegraphics[width=0.4\textwidth]{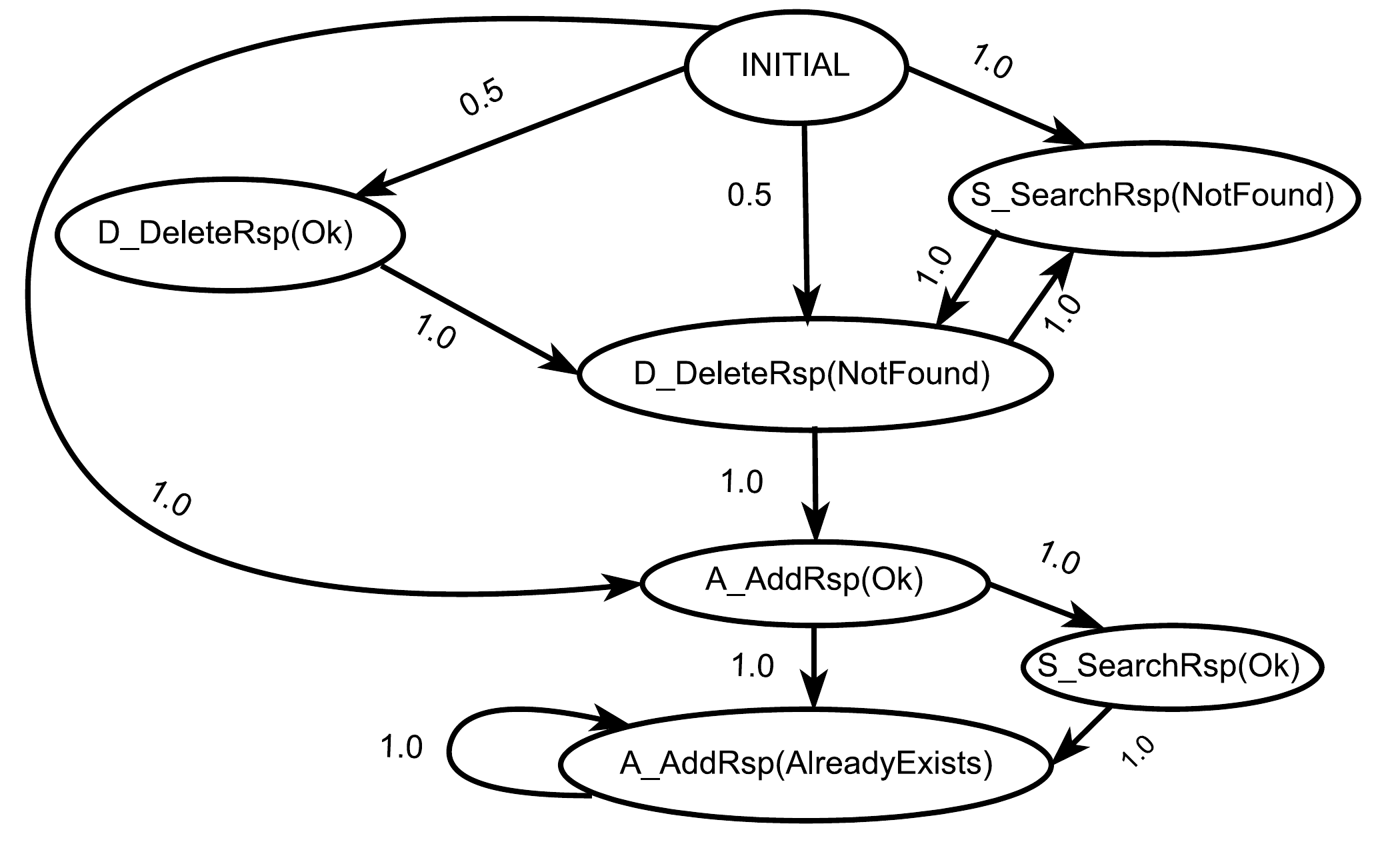}
	\caption{Inferred probabilistic message dependency model (key payload)}
	\label{fig:probabilisticModel}
\end{figure}

The probability is computed based on the ratio of the number of response messages (of a particular type) to the number of request messages in the training trace. For example, for all the \messagetype{Delete} requests that are invoked as the first request (on a particular key payload), 50\% of them lead to \messagetype{Ok} responses, and the remaining 50\% lead to \messagetype{Not Found} responses. Therefore, both edges from  \messagetype{Initial} to \messagetype{D\_DeleteRsp(Ok)} and  \messagetype{Initial} to \messagetype{D\_DeleteRsp(Not Found)} have 0.5 as their probabilities (Fig~\ref{fig:probabilisticModel}), indicating that a \messagetype{Delete} request may lead to either \messagetype{D\_DeleteRsp(Ok)} or \messagetype{D\_DeleteRsp (Not Found)} with each having the 50\% probability. 

Fig.~\ref{fig:probabilisticModel} illustrates the probabilistic message dependency sub-model (key payload) inferred from the same sequence of interactions presented in Table \ref{tab:exampleinteractionlibrary}, but with a different initial state for the key payload \dquotes{\messagetype{Judith}} (\ie, the record with the key payload \dquotes{\messagetype{Judith}} was added before the start of data collection).

In this non-deterministic or probabilistic  message dependency model, the response type for an incoming request message can be determined at runtime (as detailed in Section \ref{subsec:responseTypeSelection}) in two distinct ways: (1) randomly, or (2) based on the probabilities linked to the transitions.  Response generation experiments have been conducted using both methods for selecting the response type from the inferred model and are discussed in the next section.



\section{Evaluation}
\label{sec:evaluation}

This section presents the experimental results that assess the effectiveness of our approach. First discussed are the datasets, the evaluation methodology and criteria, as well as the techniques used for comparison. Then, the experiment outcomes are analyzed.

\subsection{Datasets}
\label{subsec:datasets}
Our evaluation experiments use service interaction traces (or datasets) gathered from both stateful and stateless services. They include CA IM (LDAP) and Bank (SOAP) for the former and Twitter (REST) and GoogleBooks (REST) for the latter. While our primary goal is to enhance the response generation accuracy for stateful services, stateless services are also considered to evaluate the approach's applicability in those contexts. The classification of service as stateful or stateless in our experiments is based on the collected interaction traces. A service that is generally stateful may be treated as stateless depending on the recorded service interactions. For instance, although GoogleBooks is a stateful (REST) web service, the trace collected for the GoogleBooks dataset exclusively includes retrieval requests (searches) and does not feature any requests to add or delete books. As a result, the GoogleBooks dataset is classified as a stateless service dataset since it reflects only the stateless capabilities of the service.

\smallskip
\noindent {\bf CA IM (LDAP):}
The Lightweight Directory Access Protocol (LDAP) is commonly used for accessing and maintaining distributed directory information services over the Internet Protocol \cite{yeong:1995}. It is a binary protocol that uses the ASN.1 encoding to encode and decode text-based message information to/from its binary representation \cite{bapat:1994}. The CA IM \cite{caidentitymanager} enterprise directory service implementing LDAP was used to obtain the CA IM (LDAP) datasets (henceforth also referred to as the LDAP datasets). The textual representation of LDAP interaction traces are used, containing eight types of LDAP request messages: \messagetype{Bind}, \messagetype{Add}, \messagetype{Delete}, \messagetype{Search}, \messagetype{Modify DN}, \messagetype{Modify}, \messagetype{Compare}, and \messagetype{Unbind}. The two LDAP datasets contain \numprint{20,157} and \numprint{19,930} interactions for the \textit{clean start} and \textit{non-clean start} scenarios, respectively (see below).

\smallskip
\noindent {\bf Bank (SOAP):}
The Simple Object Access Protocol (SOAP) is a lightweight protocol designed for the exchange of structured information in a decentralized, distributed environment~\cite{box:2000}, commonly utilized by Web Services. SOAP messages leverage XML technologies to establish an extensible messaging framework, providing a message structure that can be transmitted over various underlying protocols. The Bank (SOAP) datasets (hereafter also referred to as the SOAP datasets) are derived from a Banking web service and encompass interactions involving five request types: \messagetype{createNewAccount}, \messagetype{deposit}, \messagetype{withdraw}, \messagetype{getAccount}, and \messagetype{closeAccount}. Each of the two SOAP datasets comprises \numprint{20,000} interactions for both the \textit{clean start} and \textit{non-clean start} scenarios.

\smallskip
\noindent {\bf Twitter (REST):}
The Twitter (REST) API \cite{twiiter:2014} provides Web application developers with a number of services to facilitate the programmatic use of Twitter functionalities. The Twitter service is used here in a stateless capability and its data/state changes are not observable in the interaction messages. As such, only one Twitter dataset is used for both the \textit{clean start} and \textit{non-clean start} scenarios. 
It contains \numprint{1,465} interactions of six request types: \messagetype{StatusesShow}, \messagetype{StatusesUpdate}, \messagetype{StatusesUserTimelineJsonUserid}, \messagetype{SearchTweets}, \messagetype{StatusesUserTimelineJsonScreenname}, and
\messagetype{FriendshipsShow}.

\smallskip
\noindent {\bf GoogleBooks (REST):}
GoogleBooks is a REST service offered by Google Inc. for searching full-text books and retrieving book-related information \cite{googlebooks:2018}. Like the Twitter service, the GoogleBooks dataset is collected when it operates solely in a stateless capacity. As a result, there is only one GoogleBooks dataset that covers both the \textit{clean start} and \textit{non-clean start} scenarios. This dataset consists of \numprint{1,913} interactions involving three request types: \messagetype{searching} for a book volume, \messagetype{retrieving} a specific book volume, and \messagetype{retrieving} information about public bookshelves.

\smallskip
\noindent 
Each of the above datasets or interaction traces used in our experiments is collected by intercepting the communications between a client system (representing the SUT) and the target service. The client sends requests to the service according to various use scenarios of the service, and the client requests and service responses (\textit{i.e.}, the TCP/IP packets) are intercepted and recorded using Wireshark \cite{combs:2007wireshark}. Further processing of these recorded raw messages are conducted according to our own prior work in~\cite{hossain2022extracting} and~\cite{jiang:2019positional}, including message format identification. For model inference and analysis, the main external tool used is the ktail algorithm \cite{biermann:1972} as discussed in previous sections; otherwise, all other tools are developed according to our own algorithms as reported in this paper.

\subsection{Evaluation Approach and Criteria}
\label{subsec:evaluationmetric}
The widely recognized \textit{cross-validation} approach \cite{devijver:1982} has been used for \emph{accuracy} evaluation, which allows each interaction to be validated exactly once and guarantees that all interactions are included in both training and testing phases. Specifically, a 10-fold cross-validation is implemented. Since our goal is to synthesize responses for stateful services (where the sequence of interactions influences the state of the services), it is essential to track the service interactions specific to each data record separately from other interactions. Consequently, the interaction trace is divided according to the \textit{key payloads} of the request messages, and the folds are generated based on the number of data record-based partitions. As a result, not every fold contains the same number of interactions in the training and testing datasets. Nonetheless, our technique ensures that all interactions are utilized in both testing and training.

To measure the \emph{accuracy} of response generation, the synthesized responses are compared with the actual recorded responses in the collected interaction trace, where the latter are considered to be the ground truth. In doing so, the following criteria are used to quantify the degree of accuracy in the generated responses: 

\begin{enumerate}[leftmargin=*]
\item 
\textbf{Identical:} The synthesized response is identical to the expected response (\ie, the recorded response in the trace, including all payload fields (see Example 1 in Table \ref{tab:exampleaccuracycriteria}).
\item 
\textbf{Data consistent:} The synthesized response has the expected response type and the expected payloads in the \emph{critical} payload fields (\ie, the key payload fields and protocol-meta data fields), but may differ in other payload fields (see Example 2 in Table \ref{tab:exampleaccuracycriteria}, where the generated \dquotes{{\small\tt cn}} and \dquotes{{\small\tt id}} fields are identical to those in the expected response but the \dquotes{{\small\tt sn}} field differs).
\item 
\textbf{Protocol exact:} The synthesized response has the expected response type, but differs from (some of) the expected payload fields, including the key payload fields and protocol-meta data fields (see Example 3 in Table \ref{tab:exampleaccuracycriteria}, where the generated \dquotes{{\small\tt op}} (response type)  and \dquotes{{\small\tt  result}} (response code) are identical to the expected but the \dquotes{{\small\tt id}}, \dquotes{{\small\tt cn}} and \dquotes{{\small\tt sn}} fields differ).
\item 
\textbf{Protocol plausible:} The synthesized response is among the response types of the request but not the expected one. The generated payload fields may or may not have the expected value (see Example 4 in Table \ref{tab:exampleaccuracycriteria}).
\item 
\textbf{Well-formed:} The synthesized response has the wrong response type for the request but is of a valid response type for another request type (see Example 5 in Table \ref{tab:exampleaccuracycriteria}).
\item 
\textbf{Malformed:} The synthesized response does not meet any of the above criteria, \ie, it does npt have a valid response type for the service at all (see Example 6 in Table \ref{tab:exampleaccuracycriteria}).

\end{enumerate}

\begin{table}[!t]
    \centering
    \caption{Accuracy criteria examples for synthesized responses}
    \label{tab:exampleaccuracycriteria}
    \includegraphics[width=.48\textwidth]{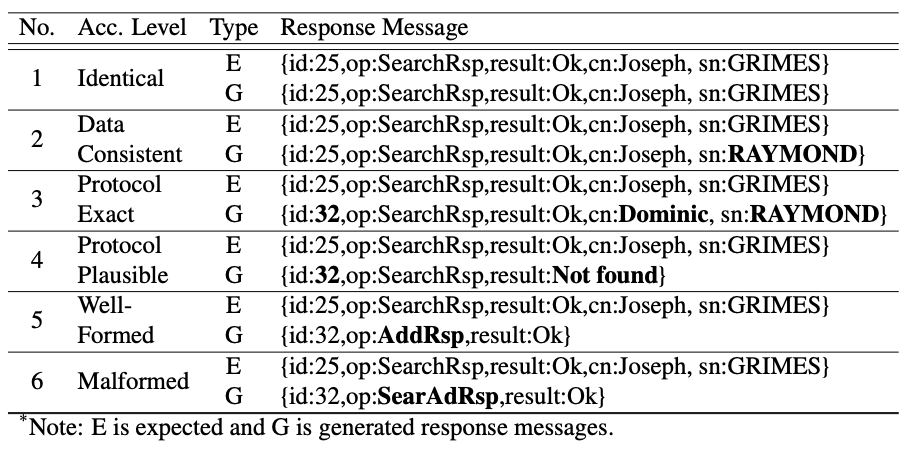}
\end{table}

In addition to the accuracy of the synthesized response message, the \emph{efficiency} of response generation is also measured in terms of the time taken to produce a response and the maximum memory required during the execution of the experiments.

\subsection{Compared Techniques}
Our approach is evaluated against three state-of-the-art existing service virtualization techniques for synthesizing response messages.
The first technique is the \textit{Whole-Cluster} OSV approach~\cite{du:2013}, which identifies the closest matching interaction from the recorded interactions for the incoming request and synthesizes a response by substituting the relevant payloads.
The second method compared is the Message Prototype-based OSV approach \cite{versteeg:2016}. This method employs multiple sequence alignment to create message prototypes for various request types. At runtime, the incoming request message is compared to these generated prototypes instead of the recorded interactions.
The third technique is the Sequence-to-Sequence Based Virtualization (SSBV) approach \cite{enicser2018testing}, a machine learning method for response synthesis. This technique involves training a basic Long Short-Term Memory (LSTM) Seq2Seq model using the training dataset, processing the input message character by character to generate the response in a similar fashion.

\subsection{Experiment Results}
\label{subsec:evaluationResultChapter6}
Our approach is evaluated from the perspectives of \textit{accuracy} and \textit{efficiency}, addressing the following questions:

\smallskip
\noindent
\textbf{Q1 (Accuracy):} Can the response generation method outlined in Section \ref{subsubsec:responseGeneration} produce more accurate responses for both stateless and stateful services using the inferred service behavior model described in Section \ref{subsec:serviceModelInference}? Do the inferred message formats detailed in Section \ref{subsec:traceAnalysis} enhance the accuracy of the generated responses?

\smallskip
\noindent
\textbf{Q2 (Efficiency):} Is our method capable of synthesizing responses in a reasonable timeframe while using achievable resources (compared to the actual services)?

\subsubsection{Accuracy}
This section discusses the experimental results concerning the accuracy of our approach and compares them to three existing techniques for response synthesis. The experiments are conducted on traces collected under two distinct scenarios: (i) Clean start, where the data records used for collecting traces were not added before the commencement of data collection; (ii) Non-clean start, where the data records related to the captured interactions in the traces may or may not have been added (\ie, some data records have been added while others were not) before data collection began.

\setlength{\belowcaptionskip}{-5pt}

\begin{figure}[!t]
	\centering
	\includegraphics[width=1\columnwidth]{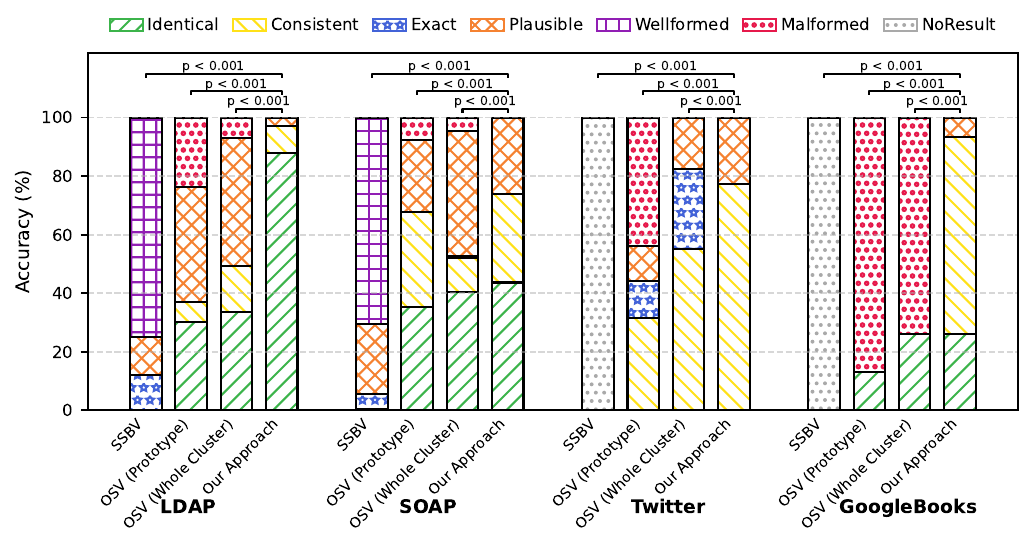}
	\caption{Response generation result (clean start). \textit{p}-values were computed for each dataset by comparing the outcome distributions of each state-of-the-art approach with our approach across all response categories, using Pearson's chi-squared test of independence.}
	\label{fig:responseGenerationResult(cleanStart)}
\end{figure}

\smallskip

\noindent {\bf Clean Start:} 
Fig.~\ref{fig:responseGenerationResult(cleanStart)} illustrates the accuracy of our approach compared to the existing methods in generating responses. Our approach consistently outperforms the three existing methods across all response categories for all four datasets, with statistically significant improvements observed ($p<0.001$).
Our approach surpasses the others in producing accurate responses for stateful services such as CA IM (LDAP) and Bank (SOAP). 

\textbf{CA IM (LDAP) dataset:} For the LDAP dataset, our method achieves 87.98\% (95\% confidence interval (CI): 87.47\% to 88.49\%) \emph{identical} responses, whereas the SSBV, Prototype OSV, and Whole-Cluster OSV yield only 0.11\%, 30\%, and 33\% identical responses, respectively. This demonstrates that our approach more than doubles the accuracy of the existing OSV methods (Whole-Cluster and Prototype) in generating \textit{identical} responses. The SSBV, on the other hand, produces a minimal number of identical responses (\ie, less than 1\%).

At the next high level of accuracy, all four approaches produce \emph{data-consistent} responses for the LDAP dataset. 
As none of the methods accurately account for a proper \emph{data model} for the \textit{record-specific} payload fields (such as \messagetype{mobile number} and \messagetype{address}), varied payloads are generated for those fields. 

Moving down the accuracy scale, only SSBV generates 
\emph{protocol-exact} responses for the LDAP dataset. This suggests that some of the responses produced by the SSBV lack accurate payloads in the \emph{critical fields}, such as the message ID.

At the lower end of the accuracy scale, our approach generates very few \emph{protocol-plausible} responses only 2.92\% (95\% CI: 2.72\% to 3.12\%), whereas the SSBV, Prototype OSV, and Whole-Cluster OSV generate 13\%, 39\%, and 43\% protocol-plausible responses, respectively. This is due to the fact that our inferred message dependency model keeps track of the service state for each data record (key payload) and the large majority of the generated responses have the correct response type. The small number of responses that have the wrong type (protocol-plausible) is due to the reason that LDAP allows the key payload to be changed by the \messagetype{ModifyDN} request. As such, our inferred model is unable to keep track of the service state for the data record with the changed key payload, and responses of the wrong type are generated for future requests on the updated key payload. However, they are a very small percentage.

Crucially, our approach does not produce any \emph{malformed} responses, whereas the Whole-Cluster and Prototype OSV approaches generate malformed responses at rates of 7\% and 23\%, respectively. Both of these approaches fail to take the message structure into account during payload substitution, which leads to the generation of malformed responses. In contrast, our approach infers message formats and uses them to determine payloads for the responses, ensuring that no malformed responses are generated. SSBV does not create malformed responses either, as it selects one response from the training dataset and utilizes that same response for all incoming requests. This results in 75\% \emph{well-formed}  responses for the LDAP dataset, though it achieves significantly lower percentages at higher accuracy levels.

\textbf{Bank (SOAP) dataset:} For the SOAP dataset, our approach achieves 43.67\% (95\% CI: 42.98\% to 44.36\%) identical responses, while the SSBV, Prototype OSV, and Whole-Cluster OSV yield 0.51\%, 35\%, and 40\% identical responses, respectively. This indicates that our approach still generates a higher number of responses at greater accuracy levels, although not to the same extent as with the LDAP dataset. Specifically, our approach produces 30.28\% (95\% CI: 29.65\% to 30.91\%) data-consistent and 26.05\% (95\% CI: 25.44\% to 26.66\%) protocol-plausible responses for the SOAP dataset, higher than for the LDAP dataset. 

The increased number of data-consistent responses for the SOAP dataset can be attributed to a number of reasons. First, responses for most request messages in SOAP, such as \messagetype{getAccount}, \messagetype{withdraw} and \messagetype{deposit}, include the non-key payload field \messagetype{balance}, which usually changes from one request to the next. Since our approach currently does not account for how payloads evolve over time, the generated responses lack the exact values for the \messagetype{balance} field.

The reason for generating more protocol-plausible responses for SOAP than for LDAP is twofold. First, the LDAP Directory Service operates as a CRUD (Create, Read, Update, and Delete) service, allowing a data record to be deleted via a \messagetype{delete} request and subsequently re-created with an \messagetype{add} request. In contrast, a bank account in the SOAP Bank Service can only be opened once, and there are no requests in the SOAP trace to reopen an account after it has been closed with a \messagetype{closeAccount} request.

Second, the types of LDAP responses (e.g., \messagetype{DeleteRsp (Success)} and \messagetype{DeleteRsp(Not Found)}) for incoming requests depend solely on the key payload. Therefore, different response types are generated based on whether the data record is present, deleted, or absent in the directory. On the other hand, the response types (e.g., \messagetype{withdrawResponse (Success)} and \messagetype{withdrawResponse (Fail)}) for a \messagetype{withdraw} request in the Bank (SOAP) service depend not only on the key payload (i.e., the account number) but also on the current balance and the amount requested for withdrawal. Consequently, a \messagetype{withdraw} request could yield a \messagetype{withdrawResponse(Fail)} response due to either an invalid account number or insufficient funds, making it challenging for our approach to track changes in the service state for such non-key payloads.

\textbf{Twitter and GoogleBooks datasets:} For stateless services, specifically Twitter and GoogleBooks, the SSBV technique fails to generate any responses. The response messages for these datasets are significantly lengthy, with the maximum lengths reaching \numprint{90,295} characters for Twitter and \numprint{134,376} characters for GoogleBooks. The SSBV technique requires the creation of three \messagetype{NumPy} arrays for one-hot encoding of the input messages, each consuming approximately \numprint{5} GB of memory. Additionally, Long Short-Term Memory (LSTM) networks maintain an intermediate state at each step, which further increases memory requirements during model training. As a result, the training process was automatically terminated due to an out-of-memory error, even with a memory allocation of \numprint{200} GB. This indicates that the SSBV technique is unsuitable for generating responses when dealing with lengthy messages.

Among the three remaining approaches, our method does not produce any \emph{malformed} responses, whereas both the Whole-Cluster and Prototype OSV methods do, for the same reasons discussed earlier. None of the approaches generates \emph{identical} responses for the Twitter dataset, primarily due to the presence of large or frequently changing payloads such as timestamps, tweet IDs, user IDs, or engagement metrics such as likes and retweets. However, our method achieves a \emph{data-consistent} rate of 77.2\% (95\% CI: 75\% to 79.4\%), while the Prototype OSV and Whole-Cluster OSV methods yield 31\% and 55\% data-consistent responses, respectively. As illustrated in Fig.~\ref{fig:responseGenerationResult(cleanStart)}, all three approaches produce \emph{protocol-plausible} responses for the Twitter dataset. This occurs because certain requests can result in different types of responses, yet the underlying reasons for these variations are not discernible from the interaction trace alone. For instance, the response to a \messagetype{searchTweets} request may be empty or include several tweets, depending on the search query. But it is impossible to determine the cause of this variability by merely analyzing the interaction trace. 

In the case of the GoogleBooks dataset, all three techniques do generate some \emph{identical} responses. This is attributed to the \messagetype{search\_bookshelf} responses, which return an \dquotes{error} message when the \messagetype{bookshelf\_id} is not found, and these \dquotes{error} responses share the same message structure. Similar to the Twitter dataset, our approach also generates protocol-plausible responses for the GoogleBooks dataset because the same request can produce different response types at different times, yet the reasons behind these variations are not captured in the interaction trace.

Overall, our approach surpasses existing service virtualization techniques in generating accurate responses for both stateful and stateless services. It achieves notably higher percentages of identical and data-consistent responses, with no malformed responses. The message dependency model aids in accurately selecting the appropriate response types for incoming requests, while the equality-based substitution rules ensure (some) correct payloads are inserted into the synthesized responses. Unlike the OSV approaches, our approach effectively avoids generating malformed responses due to the use of inferred message formats in crafting the responses.

\smallskip

\noindent {\bf Non-clean Start:}
Our approach infers a probabilistic message dependency model for scenarios with a non-clean start, where some data records are already present in the service before data collection begins. In these situations, the response types for an incoming request can be determined in two ways: (i) randomly, or (ii) probabilistically based on their frequency of occurrence in the training dataset.

\begin{figure}[!t]
	\centering
	\includegraphics[width=1\columnwidth]{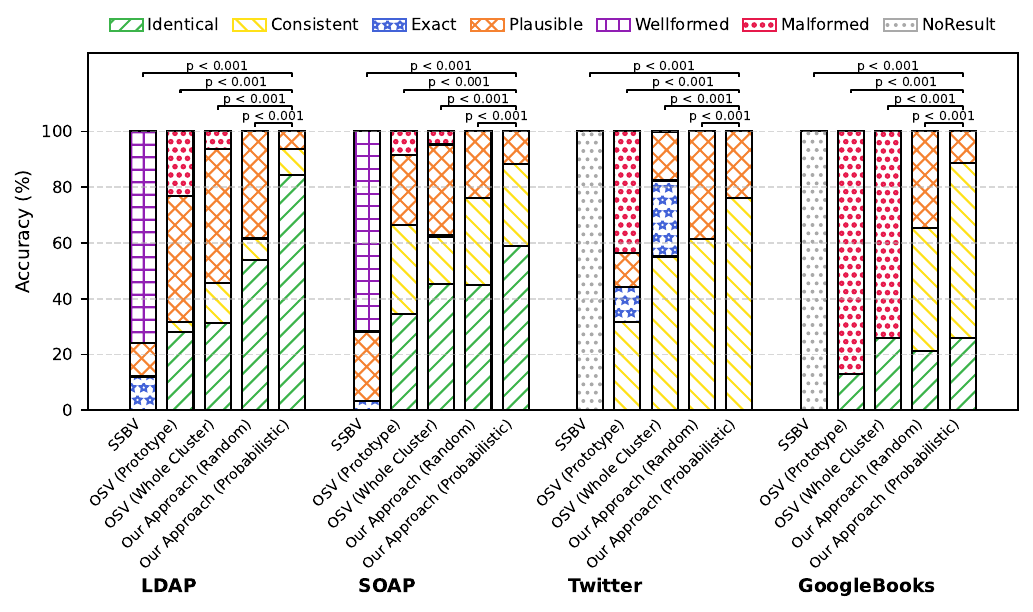}
	\caption{Response Generation Result (Non-clean Start). \textit{p}-values were computed for each dataset by comparing the outcome distributions of each approach with our probabilistic-based approach across all response categories, using Pearson's chi-squared test of independence.}
	\label{fig:responseGenerationResult(probabilistic)}
\end{figure}

Fig.~\ref{fig:responseGenerationResult(probabilistic)} shows the accuracy of response synthesis for the non-clean start scenario. Our probabilistic approach consistently outperforms the random variant of our method as well as existing baseline approaches, with statistically significant improvements ($p < 0.001$). For identical responses, the probabilistic method achieves 84.21\% (95\% CI: 82.0\% to 86.4\%), while the random approach attains 53.90\% (95\% CI: 49.0\% to 58.8\%), both surpassing the accuracy of existing baselines. Similarly, for data-consistent responses, the probabilistic and random approaches achieve 9.36\% (95\% CI: 8.1\% to 10.7\%) and 7.69\% (95\% CI: 5.5\% to 9.8\%) accuracy respectively, outperforming prior methods. These results demonstrate that both variants of our approach significantly improve the generation of identical and data-consistent responses for stateful services (\ie, LDAP and SOAP), with the probabilistic approach showing a clear advantage in accuracy and precision.

The improvement in generating identical responses for the SOAP dataset is less pronounced than for LDAP, due to similar factors observed in the clean start scenario. Notably, when selecting the type of response based on probabilistic scores, our approach achieves better accuracy across all datasets compared to randomly selecting response types from the message dependency model. This improvement arises because the probabilistic method takes into account the ratios of response occurrences in the training dataset when determining response types for incoming requests.

For stateless services, the same datasets as in the clean start scenario are used as changes in service state are not observable in the interaction trace. Consequently, our approach using the probabilistic message dependency model achieves similar accuracy to the deterministic model used in the clean start scenarios for the Twitter and GoogleBooks datasets.

\smallskip

\noindent {\bf Summary:}
In general, our approach consistently outperforms the benchmark techniques used for comparison in synthesizing responses for both stateful and stateless services. The accuracy of the synthesized responses is evaluated based on the criteria outlined in Section \ref{subsec:evaluationmetric}, where identical responses represent the highest degree of accuracy (\ie, the generated responses match the expected ones exactly) and malformed responses indicate the lowest (\ie, the generated responses do not adhere to any valid message format).

As demonstrated in Figs~\ref{fig:responseGenerationResult(cleanStart)} and \ref{fig:responseGenerationResult(probabilistic)}, our approach excels in both clean and non-clean start scenarios. Notably, our approach does not produce any malformed responses, as it effectively utilizes message formats during response formulation. In contrast, existing techniques, such as the Whole-Cluster and Prototype OSV approaches, do not account for message structure, leading to the generation of malformed responses.
Additionally, our inferred message dependency model and payload substitution rules enhance the generation of identical and data-consistent responses, especially for stateful services. 


In the interaction traces from stateless services (\ie, Twitter and GoogleBooks), the change in service state is not observable, and the responses include substantial amounts of application data payloads that are not present in the request messages. Despite these challenges, our approach continues to outperform existing techniques by generating a higher number of identical and data-consistent responses.

\subsubsection{Efficiency}
The efficiency assessment of our approach focuses on the response generation time and memory consumption. The response generation code has been instrumented to record the response generation times and memory usage across all datasets. The experiments were carried out on the OzStar supercomputer, which features 10 x86 Intel Skylake Xeon Gold 6140 CPUs operating at 2.3 GHz, with each CPU having 20 GB of memory.
For consistency in comparison, our approach and the compared techniques (\ie, Whole-Cluster OSV, Prototype OSV, and SSBV) were run on the same CPUs but at different time intervals. Furthermore, the  response times of the actual services are also collected during interaction trace collection. 

Below is a brief summary of the experimental results (due to space limitation); full details can be found in Supplementary Section S1.

{\bf Response Generation Time:} The experiments show that our approach significantly outperforms others in response generation time, generating a response in a fraction of a millisecond. For instance, it is 159 times faster than the actual services and up to 10,653 times faster than the Whole-Cluster OSV method.

{\bf Memory Usage:} The experiments also show that our approach is more memory-efficient during runtime, consuming less memory than other methods by utilizing pre-inferred rules instead of dynamic rule inference or extensive message comparisons. For instance, it uses 13 times less memory than the Whole-Cluster OSV method and 11 times less than the Prototype OSV method.

In summary, our approach excels in both time and memory efficiency, compared to existing methods.


\section{Limitations and Threats to Validity}
\label{sec:discussion}

In this section, we further reflect on our approach and discuss its limitations. The threats to validity concerning the evaluation of our approach are also discussed.

\smallskip
\noindent \textbf{Limitations:} One of the limitations of our approach is its inability to \emph{automatically} identify the key payload fields from the interactions on which the state of the service depends. The user has to identify the key payload field in the messages by providing a regular expression for it. The identification of the key payload fields may not always be possible for legacy or un-documented services. When the key payload fields can not be identified, our approach produces a less precise message dependency model, which restricts it from generating identical responses. However, it will still be able to generate  responses at the accuracy level of data-consistent or lower in such cases. As such, the inability to identify the key payload field does not limit the applicability of our approach in virtualizing un-documented or legacy services.

Another limitation is our approach's inability in generating exact payloads for the non-key payload fields. 
Our approach does not capture the way in which the data values of non-key payload fields change over time, and hence the generated response messages contain inexact values for those fields, \ie, some data-consistent responses are generated (instead of identical ones). For example, for LDAP, the synthesized response of a \messagetype{search} request may contain outdated values in the mobile number field if the mobile number has been updated by a \messagetype{modify} request before the \messagetype{search} request.  

Our approach has solely relied on the captured interaction trace for deriving a service model. As a result, the derived service model simulates only those service characteristics/behaviors observed in the interaction trace. Consequently, the derived model does not imitate any characteristics not present in the collected interaction trace. However, the impact of this limitation can be minimized by collecting the interaction trace covering as many typical/representative use scenarios as possible.

Furthermore, our approach assumes that the communication between the service and its client is synchronous. It does not consider asynchronous communication between a service and its client. 
Finally, our approach has been designed primarily for services using a textual protocol. It is still applicable to services using a binary protocol as long as a suitable decoder/encoder is available to translate the messages between the binary form and a textual representation. 
The issue of analyzing binary protocols will be further investigated as part of future work. 

\smallskip
\noindent \textbf{Threats To Validity:} Our approach has been evaluated on interaction traces (datasets) collected from four real-world services, and the experiments have shown that our approach achieves higher accuracy than the existing state-of-the-art approaches. However, a generalization can not be made that our approach can achieve such high accuracy for services not tested. Nevertheless, the four datasets represent widely used services adopting some of the most typical messaging protocols. Future work will further investigate the accuracy of our approach when applied to other services and messaging protocols. 

Furthermore, the inferred message dependency model is used to determine the types of response messages for incoming requests, and it does so by observing the sequences of messages that appear in the collected interaction trace. As such, the model would not be able to determine the exact response type for the unseen request types or unseen sequence of messages. This may happen for non-representative interaction traces (datasets). However, the generated responses may still be useful for the scenarios where the exact response is not required (\eg, for performance testing). Most importantly, our approach generates accurate responses for the datasets that represent the actual characteristics of the services.


\section{Conclusions}
\label{sec:conclusion}
In this paper, we have presented a new approach to service virtualization by creating model-based emulations for stateful services that can be used in testing environments for highly connected enterprise systems. Our approach synthesizes the dependency relationships among service messages by mining the contextual information embedded in a service's interaction trace, and provides a service behavior model that can be used to track the service state changes at run-time when the service interacts with a client or system under test (on behalf of the actual service). 
Our approach further utilizes the inferred message formats when formulating response messages to ensure their structures are valid. It also considers the equality relationships between message payload fields in populating payload values for the generated response messages. 

The experiments with four real-world services have demonstrated that our approach can achieve higher accuracy in generating response messages, especially for stateful services, with adequate performance and reasonable memory consumption when compared to the actual services and existing approaches.

In future work, we aim to enhance our model inference technique to automatically identify the key payload fields of a service from its interaction trace. Further relationships (beyond equality) among the message payload fields of a service and their changes over time will also be explored, to determine the exact payloads when generating response messages for the service. Another issue for further investigation is how to derive the message dependency model for services supporting asynchronous communications. 

\section*{Declaration of Competing Interest}
The authors declare that they have no known competing financial interests or personal relationships that could have appeared to influence the work reported in this paper.

\section*{Acknowledgments}
This project has received funding support from the Australian Research Council and CA Technologies.

\section*{Data Availability}
Data will be made available on request.

\section*{List of Acronyms}

\begin{tabular}{ l  l } 
 CA IM & CA Identity Manager \\
 CBV & Classification-Based Virtualization \\
 CRUD & Create, Read, Update, Delete \\
 EFSA & Extended Finite State Automata \\
 ES & Enterprise Systems \\
 ESS & Enterprise Software Systems \\
 FSA & Finite State Automata \\
 HTTP & Hypertext Transfer Protocol \\ 
 ICMP & Internet Control Message Protocol \\
 LDAP & Lightweight Directory Access Protocol \\
 LSTM & Long Short-Term Memory \\
 NTP & Network Time Protocol \\
 OSP & Opaque Service Virtualization \\
 POP3 & Post Office Protocol 3 \\
 PTA & Prefix Tree Acceptor \\
 PTP & Precision Time Protocol \\
 REST & REpresentational State Transfer \\
 RPC & Remote Procedure Call \\
 SMB & Server Message Block (protocol) \\
 SMTP & Simple Mail Transfer Protocol \\
 SOAP & Smple Object Access Protocol \\
 SSBV & Sequence-to-Sequence Based Virtualization \\
 SUT & System Under Test \\
 TCP/IP & Transmission Control Protocol/Internet Protocol \\
 

\end{tabular}




 \bibliographystyle{elsarticle-num} 
 \bibliography{references}
\end{document}